\documentclass[aps,prl,twocolumn,superscriptaddress]{revtex4-1}
\usepackage{graphicx}
\usepackage{physics}
\usepackage{multirow}
\usepackage[version=4]{mhchem}
\begin{document}

\title{Absence of strong localization at low conductivity in the topological surface state of low disorder \ce{Sb2Te3}}

\author{Ilan T. Rosen}
\affiliation{Department of Applied Physics, Stanford University, Stanford, California 94305, USA}
\affiliation{Stanford Institute for Materials and Energy Sciences, SLAC National Accelerator Laboratory, Menlo Park, California 94025, USA}
\author{Indra Yudhistira}
\affiliation{Department of Physics, National University of Singapore, Singapore 117551, Singapore}
\affiliation{Centre for Advanced 2D Materials and Graphene Research Centre, National University of Singapore, Singapore 117546, Singapore}
\author{Girish Sharma}
\affiliation{Department of Physics, National University of Singapore, Singapore 117551, Singapore}
\affiliation{Centre for Advanced 2D Materials and Graphene Research Centre, National University of Singapore, Singapore 117546, Singapore}
\author{Maryam Salehi}
\affiliation{Department of Materials Science and Engineering, Rutgers, the State University of New Jersey, Piscataway, New Jersey 08854, USA}
\author{M.~A. Kastner}
\affiliation{Department of Physics, Stanford University, Stanford, California 94305, USA}
\affiliation{Stanford Institute for Materials and Energy Sciences, SLAC National Accelerator Laboratory, Menlo Park, California 94025, USA}
\affiliation{Department of Physics, MIT, Cambridge, Massachusetts 02139, USA}
\affiliation{Science Philanthropy Alliance, 480 S. California Ave, Palo Alto, California 94306, USA}
\author{Seongshik Oh}
\affiliation{Department of Physics and Astronomy, Rutgers, the State University of New Jersey, Piscataway, New Jersey 08854, USA}
\author{Shaffique Adam}
\affiliation{Department of Physics, National University of Singapore, Singapore 117551, Singapore}
\affiliation{Centre for Advanced 2D Materials and Graphene Research Centre, National University of Singapore, Singapore 117546, Singapore}
\affiliation{Yale-NUS College, Singapore 138614, Singapore}
\author{David Goldhaber-Gordon}
\email[E-mail: ]{goldhaber-gordon@stanford.edu}
\affiliation{Department of Physics, Stanford University, Stanford, California 94305, USA}
\affiliation{Stanford Institute for Materials and Energy Sciences, SLAC National Accelerator Laboratory, Menlo Park, California 94025, USA}

\begin{abstract}
We present low-temperature transport measurements of a gate-tunable thin film topological insulator system that features high mobility and low carrier density. Upon gate tuning to a regime around the charge neutrality point, we infer an absence of strong localization even at conductivities well below $e^2/h$, where two dimensional electron systems should conventionally scale to an insulating state. Oddly, in this regime the localization coherence peak lacks conventional temperature broadening, though its tails do change dramatically with temperature. Using a model with electron-impurity scattering, we extract values for the disorder potential and the hybridization of the top and bottom surface states.
\end{abstract}

\maketitle

\clearpage

Time-reversal invariant three-dimensional topological insulators (3D TIs) are gapped materials with inverted bulk bands. At energies within the bulk bandgap, topological surface states (TSS) are guaranteed to exist~\cite{Fu2007}. Each surface state of \ce{Bi2Se3} family materials is a single two-dimensional (2D) Dirac cone in which the in-plane spin is correlated with the wave vector~\cite{Zhang2009,Xia2009}.
Whereas topologically trivial (conventional) 2D electron systems (2DES) are strongly insulating at low carrier densities because of Anderson (strong) localization~\cite{Lee1985}, TSS are expected to be impervious to localization, even under strong disorder~\cite{Nomura2007}. As far as we know, no other time-reversal invariant two-dimensional system has a metallic single-particle description in the presence of disorder -- even in graphene, intervalley scattering due to disorder leads to localization~\cite{Chen2009a}. Intuitively, TSS should not localize because large angle scattering is prohibited without a time-reversal symmetry-breaking spin-flip: states with opposite wavevector have opposite spin.

Localization can in principle occur in 3D TI thin films. Tunneling through the thickness of a film hybridizes the top and bottom TSS, opening a surface gap $2\Delta$ around the Dirac point~\cite{Linder2009,Lu2010}. The massive Dirac fermions no longer enjoy absolute protection against localization. When hybridization is small, however, large angle scattering should still be suppressed; therefore, strong localization may be suppressed at low densities where conventional 2DES would be expected to localize. Benefiting from gate-tunability and suppressed bulk conduction~\cite{Chen2010}, thin film 3D TIs provide an opportunity to study the effect of spin texture on localization physics in two dimensions.

Careful study of these systems, however, has been hampered by various materials issues: defects push the Fermi levels of the binary V-VI topological compounds (\ce{Bi2Se3}, \ce{Bi2Te3}, and \ce{Sb2Te3}) far from the Dirac point~\cite{Zhang2011}, and epitaxial mismatches between the topological insulator and the substrate introduce additional disorder~\cite{Koirala2015}. Disorder decidedly affects the electrical conduction of TSS: the mobilities of typical thin film V-VI topological materials reach only of order 100 $\text{cm}^2/\text{Vs}$. Furthermore, while electrostatic gating can tune the Fermi level to the charge neutrality point (CNP), charged impurities obscure low density transport physics in favor of transport through charge puddles~\cite{Skinner2013,Borgwardt2016,Nandi2018}. Consequently, insulating ($\sigma_{xx} < e^2/h$) time-reversal symmetry-protected 3D TI systems have only been seen in the thinnest films, where the clean-limit hybridization gap far exceeds room temperature~\cite{Kim2011,Jiang2012a,Lang2013}.

\begin{figure*}[th]
	\includegraphics[width=0.95\textwidth]{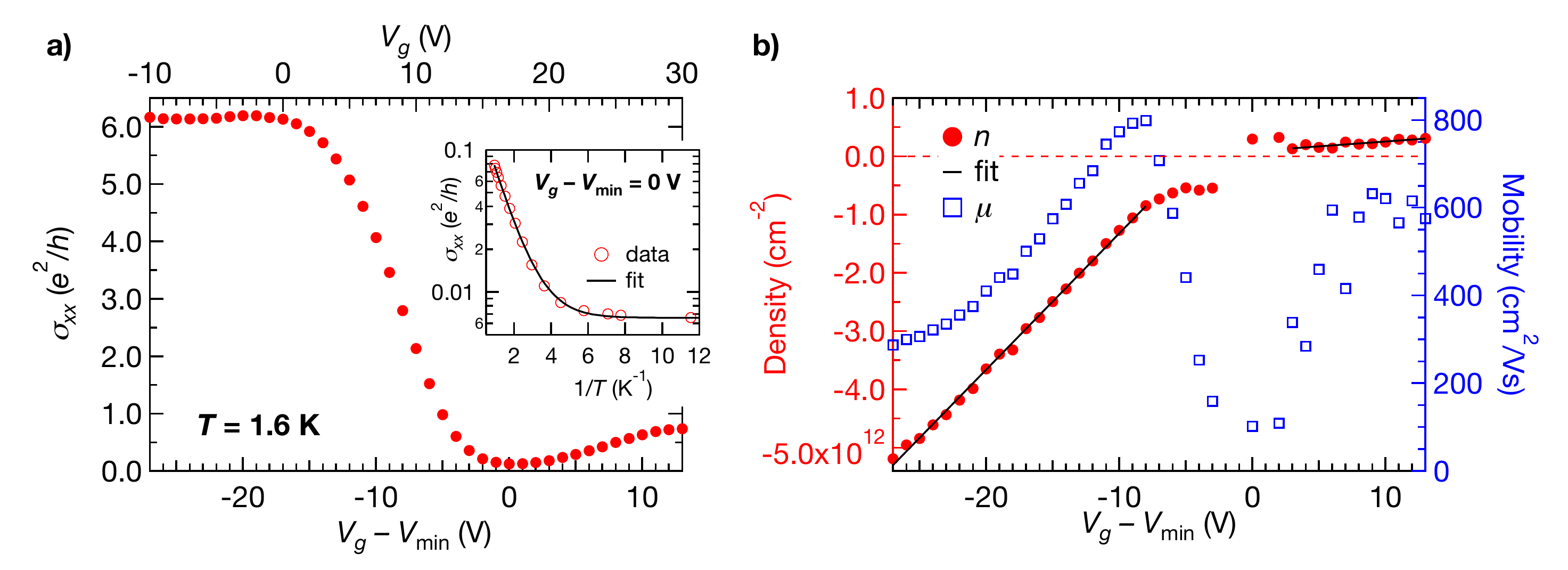}
	\caption{Transport properties. (a) Longitudinal conductivity as a function of top gate voltage at zero field and $T = 1.6 \text{ K}$. The top axis shows the actual gate voltage; the bottom axis shows the difference $\Delta V$ between the gate voltage and the conductivity minimum $V_\text{min}$. Inset: The conductivity at $V_\text{min}$ is described by an Arrhenius law plus a constant offset, which we attribute to Joule heating from the current bias. (b) Carrier density (left axis) extracted from the Hall slope, and the associated Hall mobility (right axis), shown versus $\Delta V$ at $T = 1.6\text{ K}$. Fits to the slope of the density (solid lines) yield $\alpha_h = 2.3\times 10^{11}\text{ cm}^{-2}/\text{V}$ on the hole side and $\alpha_e = 1.7\times 10^{10}\text{ cm}^{-2}/\text{V}$ on the electron side}
	\label{Fig1}
\end{figure*}

We report transport properties of a top-gated Hall bar of a novel \ce{Sb2Te3}-based thin film. The key components of this platform are 1) an epitaxially matched trivial insulator serves as a virtual substrate for the growth of the topological insulator, reducing defects, and 2) the topological insulator is counter-doped to bring its Fermi level close to the Dirac point, even before electrostatic gating. This materials platform, introduced in more detail in Ref.~\cite{Salehi2018}, offers a high mobility TSS with small and comparable surface gap and disorder potential, allowing study of the electrical transport near the Dirac point. Like other Dirac metals, this film's magnetoconductance is negative and agrees with weak anti-localization (WAL) theory at high carrier densities. Strikingly, signatures of WAL persist close to the CNP, although the conductivity $\sigma_{xx} \ll e^2/h$. This feature implies the absence of scaling to strong localization, possibly associated with the topological origin of the surface states.

\section{Methods}

\ce{Sb2Te3} was grown by molecular beam epitaxy. Interface engineering reduces disorder stemming from the lattice mismatch between topological insulator and substrate. A 15 quintuple layer (QL; 1 QL $\approx$ 1 nm) \ce{(Sb_{0.65}In_{0.35})_2Te3} (a trivial insulator) buffer layer was grown on a sapphire substrate. Growth of the topological insulator, 8 QL \ce{Sb2Te3} counter-doped by 2\% Ti, followed. A 2 QL \ce{(Sb_{0.65}In_{0.35})_2Te3} capping layer was deposited {\em in situ} to protect the TI. The thickness of the capping layer was chosen as a compromise to protect the \ce{Sb2Te3} while not impairing the efficacy of the top gate. A 50 $\mu\text{m}$ wide Hall bar was fabricated. A top gate was formed with a 40 nm alumina dielectric atop the \ce{(Sb_{0.65}In_{0.35})_2Te3} capping layer. Measurements were made in a \ce{^{3}He}/\ce{^{4}He} dilution refrigerator (30 mK to 1.2 K) and a \ce{^{4}He} cryostat with a variable temperature insert (1.5 K to 30 K) using standard lock-in techniques. To accurately measure the high resistances near the CNP at dilution refrigeration temperatures, some measurements were made using a high-impedance DC current source and a nanovoltmeter. All resistance (conductance) values are obtained through four-terminal measurements and are presented as two dimensional resistivity (conductivity).

Hall measurements at zero gate voltage and $T = 50\text{ mK}$ yielded carrier density $n=-2.5\times10^{12}\text{ cm}^{-2}$, where the negative sign indicates holes rather than electrons, and mobility $\mu = 580\text{ cm}^2/\text{Vs}$. To account for variation in the Fermi level between different cooldowns, we present gate voltage as $\Delta V_g = V_g - V_\text{min}$ where $V_\text{min}$ is the gate voltage at which the conductivity is minimized during that cooldown (between 17~V and 20~V for all cooldowns). $V_\text{min}$ is often associated with the CNP; however, given the presence of charge puddles in a disordered potential landscape, the CNP occurs precisely at $V_\text{min}$ only if electrons and holes have equal mobility.

\section{Results}

The zero-field resistance of the device is shown as a function of gate voltage in Fig.~1~(a). The carrier density, as extracted from fitting the Hall slope at applied fields $\abs{B}<0.25\text{ T}$ to a single-carrier model, is shown in Fig.~1~(b) along with the Hall mobility. At gate voltages well below $V_\text{min}$, the conductivity saturates at $\sigma_{xx}\approx 6e^2/h$, meaning the mobility $\mu\propto |n|^{-1}$. In this high-density limit, the conductivity increases logarithmically with temperature~\cite{supplement}. Moving toward $V_\text{min}$, hole carriers are depleted and the conductivity drops, reaching $\sigma_{xx} < 0.01e^2/h$ at $V_\text{min}$. As $|n|$ decreases, the temperature dependence of the conductivity evolves to an Arrhenius activation law with activation energy $\Delta_\text{Arr} = 84$~$\mu$eV at $V_g=V_\text{min}$ (Fig.~1~(a), inset)~\cite{supplement}.

The resistivity in perpendicular applied fields up to 10~T at various gate voltages is shown in the supplement~\cite{supplement}. A sharp positive quantum coherence peak in the magnetoconductance at zero field, indicative of WAL, is observed at all gate voltages. At $\Delta V_g \leq -4$~V, the coherence peak broadens with increasing temperature (Fig.~2~(a-b)), while at $\Delta V_g \geq -4$~V, the magnetoconductance flattens or switches sign as $B$ increases (Fig.~2~(c-d)). The lower the temperature, the lower the field at which the magnetoconductance switches sign. In a disordered system, aside from the coherence peak, there is a positive classical contribution to the magnetoresistance that changes from quadratic at low magnetic fields to linear at high magnetic fields~\cite{Ramakrishnan2017,Nandi2018} and saturates in some experiments~\cite{Cho2008}. A two parameter phenomenological model based on such behavior
\begin{equation}\label{eq:class}
\rho_{xx}(B)= \rho_{xx}(0)\left[1-2A+2A/\sqrt{1+(\mu B)^{2}}\right]^{-1}
\end{equation}
fits well at most gate voltages~\cite{supplement}. Here, $0\leq A\leq 0.5$ is the quadratic coefficient of magnetoreresistance i.e.~$\rho_{xx}= \rho_{xx,0}[1+A(\mu B)^2 + \cdots]$ and $\mu$ is the carrier mobility. Fig.~3 shows the coherence peaks after subtracting the background from this classical contribution. In all figures, the plotted magnetoconductance is symmetrized with respect to field~\cite{supplement}.

\begin{figure}[th]
	\includegraphics[width=0.5\textwidth]{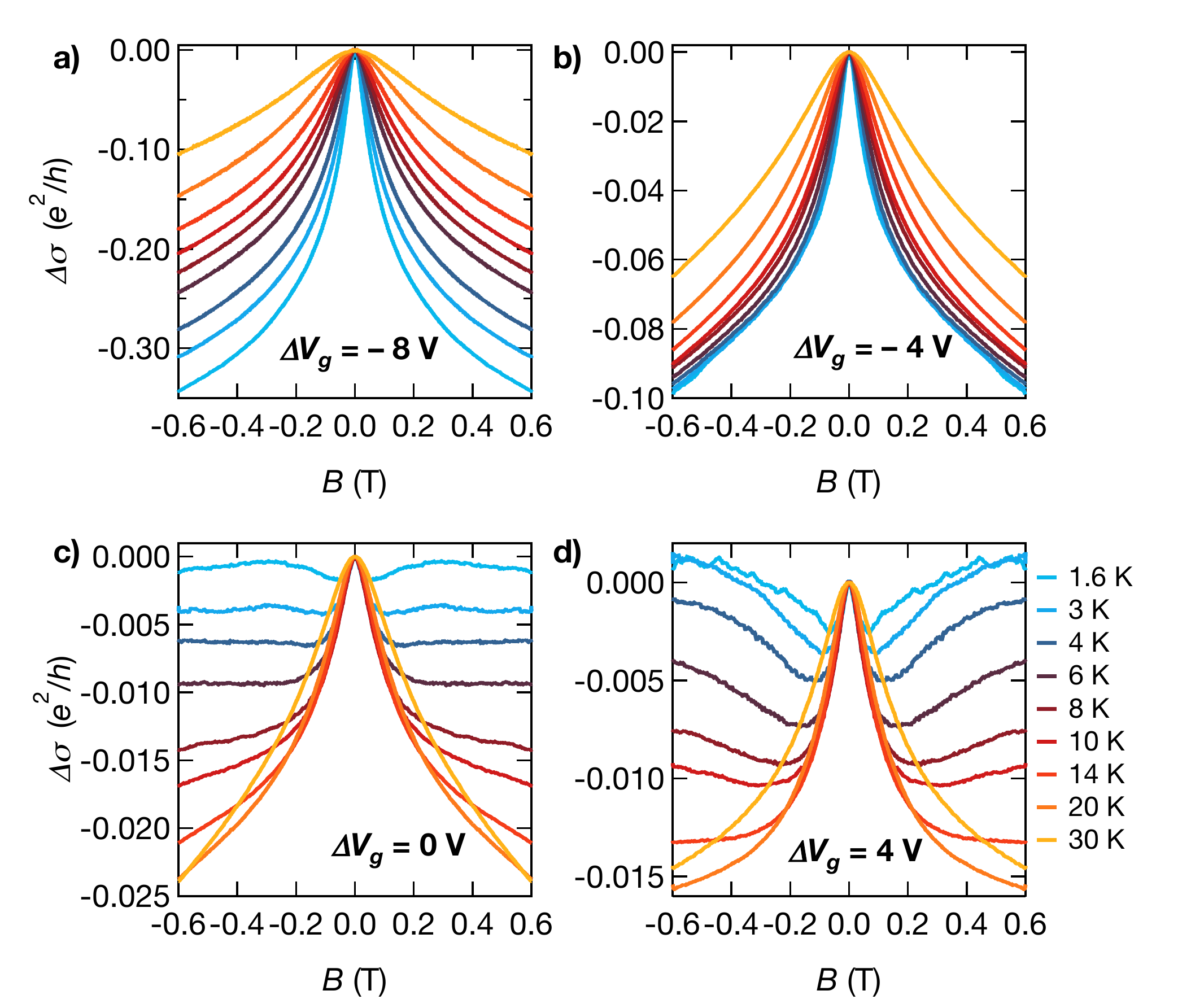}
	\caption{Magnetoconductance, indicating weak anti-localization. (a-d) Differential longitudinal conductivity, $\Delta\sigma_{xx}(T,H) \equiv \sigma_{xx}(T,H)-\sigma_{xx}(T,0)$ versus magnetic field at temperatures between $1.6\text{ K}$ and $30\text{ K}$ at $V_g-V_\text{min} = -8\text{ V}$ (a), $-4\text{ V}$ (b), $0\text{ V}$ (c), and $4\text{ V}$ (d)}
	\label{Fig2}
\end{figure}

Before proceeding, we provide quantitative estimates of the material parameters. From zero field transport and Hall coefficient ($R_\mathrm{H}$) data, we extract the density of charged impurities $n_\mathrm{imp}=(4.8\pm 2.8)\times 10^{11}\mathrm{cm}^{-2}$ lying an average distance $d=4.7~\mathrm{nm}$ from the 2DES plane, the dimensionless interaction parameter $r_s=1.3\pm0.8$, and the characteristic charge density fluctuations $n_\mathrm{rms}=(1.2\pm 0.2)\times 10^{11}\mathrm{cm}^{-2}$~\cite{supplement}.

\section{Discussion}

Arrhenius activation of the conductivity at $V_g=V_\text{min}$ confirms the presence of a surface gap. Gaps of order 100~meV have been observed through angle-resolved photoemission spectroscopy (ARPES) in 3D TI films thinner than 5 QLs~\cite{He2010,Neupane2014}. Our measured Arrhenius activation scale of 84~$\mu$eV is surprisingly small in comparison, even considering that our 8 QL film is thicker, and that $\Delta$ should decrease exponentially in film thickness. To explain the small value of $\Delta_\text{Arr}$, we note that the clean-limit surface gap $2\Delta$ could be smeared by disorder so that $\Delta_\text{Arr}\ll\Delta$~\cite{Chang2016}. Disorder smearing may also explain why many prior experiments on thin film 3D TIs do not observe a gap and instead see $\sigma>e^2/h$ at the CNP~\cite{Zhang2009a,Taychatanapat2010}.

At $\Delta V_g\lesssim -10$~V, we observed positive logarithmic temperature corrections to the conductivity. From the magnetoconductance, we know that WAL is present. WAL should contribute a negative temperature correction to conductivity:
\begin{equation}
\Delta\sigma_{xx}(T)=\frac{e^2}{\pi h}\alpha\log \left( T/T_0\right)
\end{equation}
with $\alpha = -0.5$ per channel. Our observation of positive logarithmic corrections to conductivity with increasing temperature would naively indicate weak localization (WL), with $\alpha = 1$ rather than $=-0.5$. This apparent mismatch between the signs of the temperature and magnetoconductance corrections has been previously observed and resolved by including an electron-electron interaction (EEI) contribution to the conductivity~\cite{Wang2011,Lu2014,Choi2016a}
\begin{equation}
\Delta\sigma_\text{EEI}(T)=\frac{e^2}{2\pi h}\left(2-\frac{3}{2}\tilde{F}_{\sigma}\right) \log \left( T/T_0\right),
\end{equation}
with screening factor $\tilde{F}_{\sigma}>0$~\cite{Lee1985}. The observation of overall positive temperature corrections to the conductivity means that the EEI correction dominates over the localization correction, in agreement with other experiments as well as calculations~\cite{Lu2014a}.

In principle, conventional 2DES cannot be metallic.
Surprisingly, metallic temperature dependence (higher conductivity at lower temperature, even at milliKelvin temperatures) was found in semiconductor-based 2DES of exceptional cleanliness. This is now understood to result from EEI driving the system into a metallic phase. These systems transition to insulators as carrier density is reduced. Empirically, this transition consistently occurs when the conductivity is of order $e^2/h$ in a variety of 2DES platforms including Si MOSFETs~\cite{Kravchenko1995}, GaAs/AlGaAs heterostructures~\cite{Hanein1998,Simmons1998}, graphene~\cite{Amet2013}, transition metal dichalcogenides~\cite{Radisavljevic2013}, and even other 3D TI thin films~\cite{Liao2015}.

In our system, electron-electron interactions have the opposite effect.
As discussed above, EEI causes increasing conductivity with increasing temperature.
We therefore never observe metallic temperature dependence in our device, despite the conductivity ranging from $6e^2/h$ at high carrier density to $< 0.01 e^2/h$ ($T=35\text{ mK}$) at $V_\text{min}$~\cite{supplement}.
Nevertheless, unlike conventional 2DES, 3D TIs (in the limit $\Delta=0$) are expected to have metallic single-particle descriptions. Is our system metallic? At the most fundamental level, a metal is characterized by delocalized electronic wavefunctions, not by the temperature dependance of its conductivity. Since here the temperature dependance of the resistivity fails to reflect even the presumed metallicity of the system at high doping, we must turn to the system's magnetoconductivity to address this question.

\begin{figure}[th]
	\includegraphics[width=0.5\textwidth]{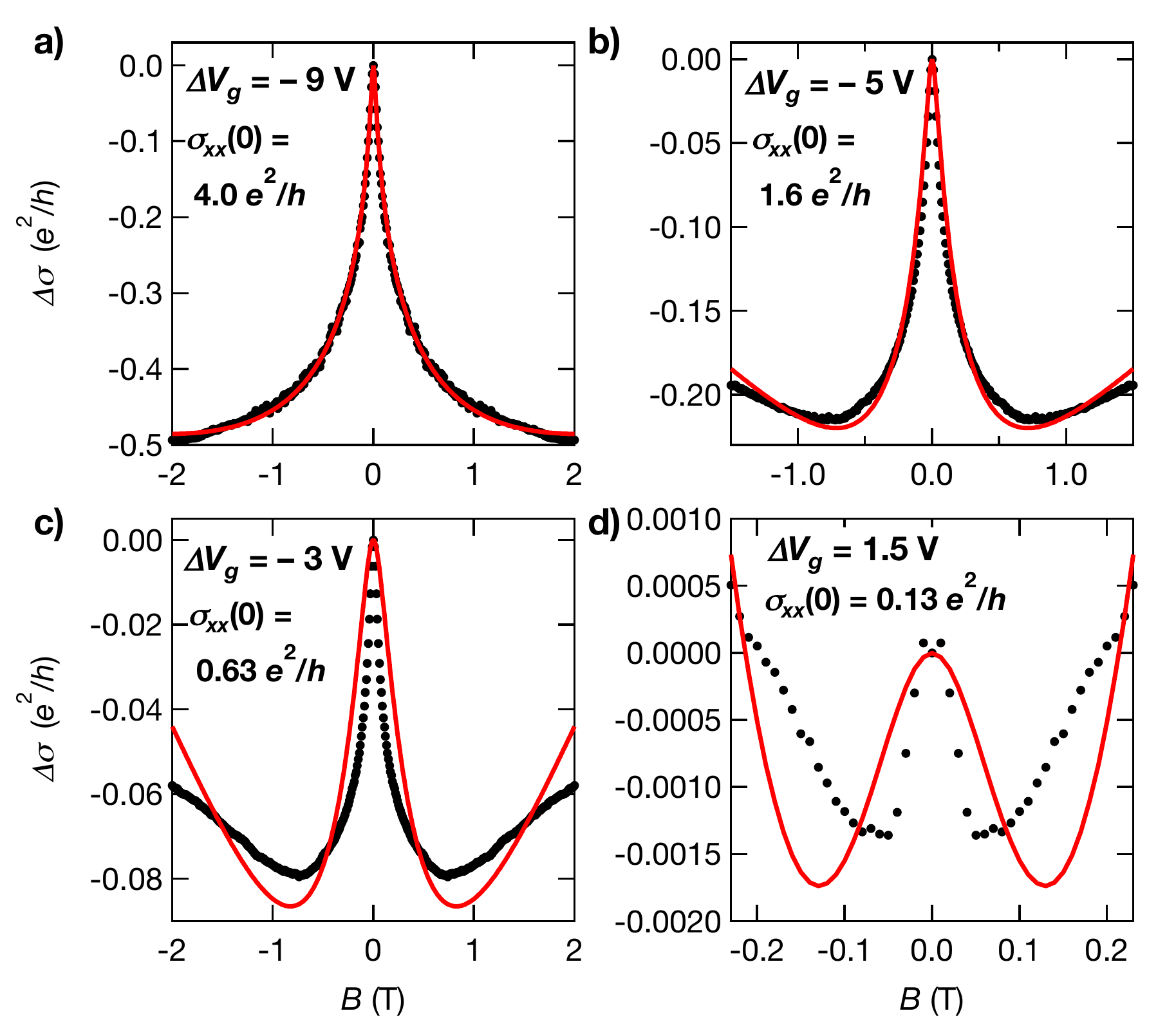}
	\caption{Crossover of the quantum transport correction from weak anti-localization to weak localization. (a-d) Differential longitudinal conductivity, $\Delta\sigma_{xx}(T,H) \equiv \sigma_{xx}(T,H)-\sigma_{xx}(T,0)$ versus magnetic field at $1.6\text{ K}$ (black dots) at $V_g-V_\text{min} = -9\text{ V}$ (a), $-5\text{ V}$ (b), $-3\text{ V}$ (c), and $1.5\text{ V}$ (d). The classical contribution to the magnetoconductance has been subtracted from the data~\cite{supplement}. The absolute conductivity at $B=0$ at each gate voltage is indicated. The HLN formula for WAL with a crossover to WL is fit to the data (red lines); the quality of fit becomes poor when $\sigma_{xx} \lesssim e^2/h$}
	\label{Fig3}
\end{figure}

In a 3D TI for which hybridization between the top and bottom surfaces is weak, electrons should exhibit weak antilocalization~\cite{Adroguer2015}. The magnetoconductance is given by the Hikami-Larkin-Nagaoka (HLN) formula~\cite{Hikami1980a}
\begin{equation}\label{eq:HLN1C}
\Delta\sigma = -\frac{e^2}{2\pi h} F\left(\frac{B}{B_\phi}\right),
\end{equation}
where $F(x) = \ln(x) + \Psi(x+\frac{1}{2})$, $\Psi$ is the digamma function, and $B_\phi=\hbar/(4el_\phi^2)$ the characteristic field associated with the electron coherence length $l_\phi$. WAL is caused by the $\pi$ Berry phase of the 2D Dirac dispersion, which suppresses backscattering. 

We observe conventional quantum transport corrections at substantial hole doping. The WAL peak broadens with increasing temperature (Fig.~2~(a)). In theory, the peak should broaden as $l_\phi\propto T^{-p}$ with $p = 0.5$ in diffusive two dimensional metals. As shown in Fig.~4, we find that $p=0.39$ at $\Delta V_g=-8$~V.

However, quantum corrections near the Dirac point differ from those at finite doping.
At gate voltages more positive than $\Delta V_g=-4$~V, the magnetoconductivity becomes non-monotonic, qualitatively departing from equation (\ref{eq:HLN1C}).
To understand this, recall that TSS hybridization in thin films should generate a Dirac mass $(\Delta)$, and the Berry phase $\phi_\textrm{b}$ should deviate from $\pi$ as $\phi_\textrm{b} = \pi \left(1-{\Delta}/{E_F}\right)$.
The Berry phase thus should induce a crossover from perfect WAL ($\phi_\textrm{b}=\pi$) in the massless (relativistic) limit to perfect WL ($\phi_\textrm{b}=0$) in the large mass (non-relativistic) regime~\cite{Lang2013,Zhang2013}, with an associated magnetic field dependence of conductivity described by a modified HLN formula~\cite{Iordanskii1994}
\begin{equation}\label{eq:HLN2C}
\begin{split}
\Delta\sigma=&-\frac{1}{2\pi}\frac{e^{2}}{h}\left[F\left(\frac{B}{B_{\phi}}\right)-2F\left(\frac{B}{B_{\phi}+B_\mathrm{\Delta}}\right)\right.\\
&\left.-F\left(\frac{B}{B_{\phi}+2B_\mathrm{\Delta}}\right)\right],
\end{split}
\end{equation}
where $B_{\phi,\Delta}=\hbar/(4el_{\phi,\Delta}^2)$ are the characteristic fields associated with the coherence length $l_\phi$ and the crossover length scale $l_\Delta$, respectively. 

The quality of fits (Fig.~3) of equation (\ref{eq:HLN2C}) is greatly improved from that of equation (\ref{eq:HLN1C}), in exchange for an additional fitting parameter. We extract $l_\Delta\sim40\mathrm{~nm}$ at all gate voltages. Unexpectedly, at the lowest field scales, we observe a WAL peak at \emph{all} gate voltages, indicating that the system does not scale to strong localization, even when $\sigma \ll e^2/h$. Furthermore, the temperature dependence of the magnetoconductance peak at $V_g\approx V_\text{min}$ is unusual, the WAL peak being more pronounced at \emph{higher} temperatures (Fig.~2).

\begin{figure}[ht]
	\includegraphics[width=0.48\textwidth]{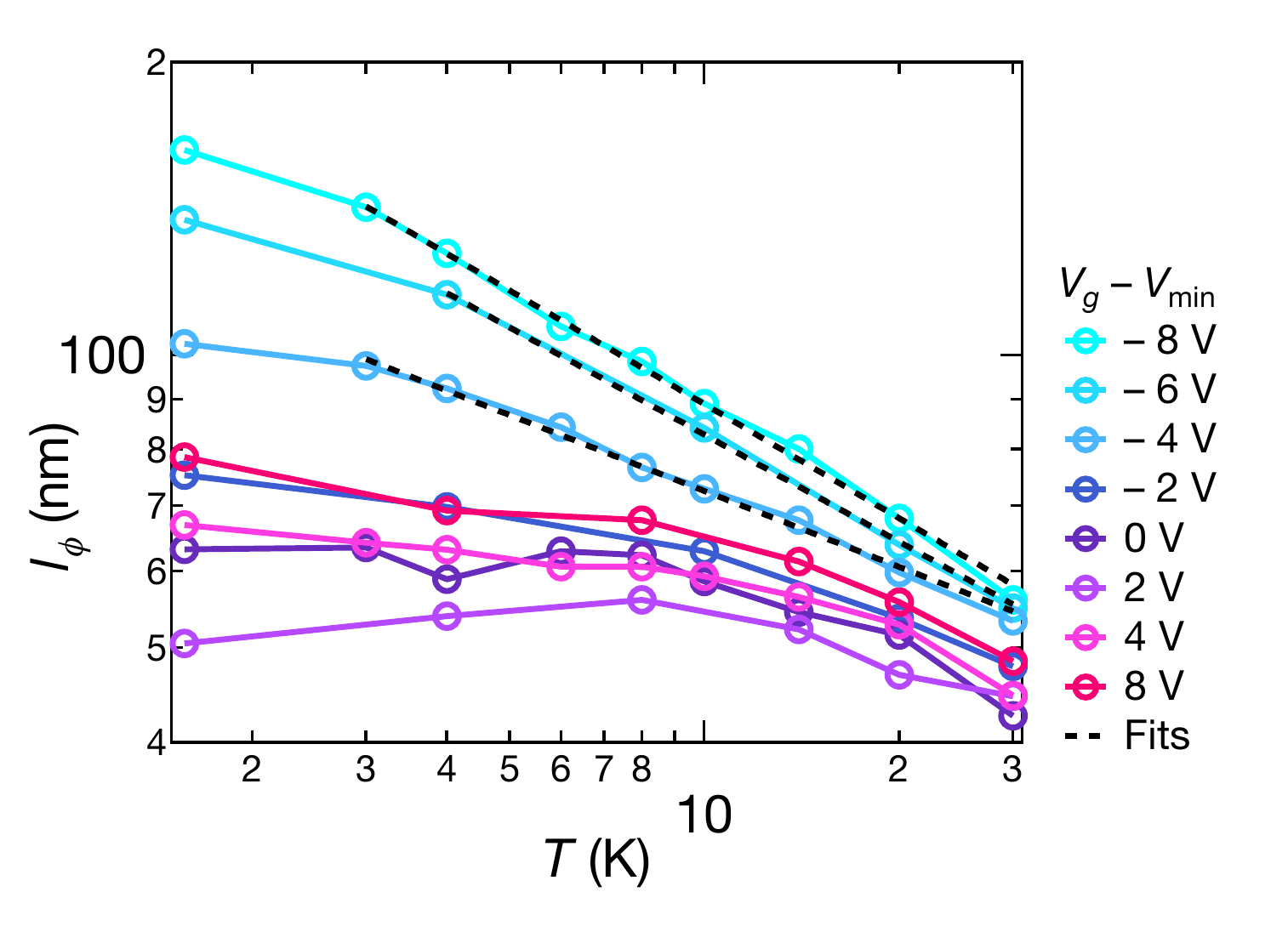}
	\caption{Coherence lengths near the Dirac point. The characteristic coherence length $l_\phi$, extracted from fitting equation (\ref{eq:HLN1C}) to the magnetoconductance, shown versus temperature at various gate voltages. A small range in $B$ is chosen when fitting to isolate the lowest order contribution. $l_\phi$ is fit by a power law in temperature $l_\phi\propto T^{-p}$, yielding $p = 0.39$, 0.37, and 0.26 at $V_g-V_\text{min} = -8$~V, $-6$~V, and $-4$~V, respectively. As $V_g$ approaches $V_\text{min}$, the temperature dependence of $l_\phi$ flattens at low temperature}
	\label{Fig4}
\end{figure}

We note that the quality of fit becomes poor near $\Delta V_g=0$, as shown in Fig.~3~(d).
Here, the data have quantum corrections only at very small magnetic fields.
We cannot make definitive statements about this observation since the quantum corrections are cleanly separable from the classical contribution to magnetoconductivity only for $\sigma \gg e^2/h$; thus, the HLN equation becomes invalid when $\sigma < e^2/h$.
However, if the data at the smallest fields are interpreted as due to quantum corrections, the extracted $l_\phi$ still decays with increasing temperature according to a power law, albeit with power roughly half that expected from EEI.  At present, we lack an explanation for this discrepancy.

We may use the extracted crossover length scale to estimate the clean-limit surface gap by dimensional analysis as $\Delta \approx \hbar v_F/l_{\Delta} = 6.7\text{ meV}$. This value is consistent with extrapolation from ARPES measurements of gaps for thinner films; as noted above, our measured transport gap $\Delta_\text{Arr}$ is much smaller, presumably because we are not in the clean limit.

An interesting pattern in the literature is that WAL is observed in topological insulators having $\sigma_{xx}>e^2/h$, while WL is observed when $\sigma_{xx} \lesssim e^2/h$~\cite{Yang2018}. This observation is explained by noting that WL in a topological insulator requires a mass gap around the Dirac point; since a gapped system insulates, we expect $\sigma_{xx} < e^2/h$.
Our results contradict this pattern: at the smallest magnetic field scales (and therefore longest length scales), we observe a quantum coherence peak with negative magnetic field corrections at all carrier densities. This signature of WAL implies delocalized electronic states. Yet, this observation holds even at low carrier densities where the longitudinal conductivity falls well below $e^2/h$. Traditionally, $\sigma_{xx}\sim e^2/h$ is associated with reaching the Ioffe-Regel criterion $k_F l\approx 1$, which predicts that metallic 2DES do not exist at lower conductivities.
Our results suggest that this device does not scale to strong localization, and instead enters an Ioffe-Regel-violating regime: a consequence of the symplectic character of the system together with disorder scattering.
In the supplement, we theoretically justify this conclusion by finding self-consistent solutions in violation of the Ioffe-Regel limit for a low-energy model of massless 2D Dirac fermions for a broader range of the dimensionless parameters $r_s$ and $n_{\text{imp}}d^2$, using a combination of analytic and numeric results.

\begin{acknowledgments}

The authors thank Eli J. Fox, Hassan Shapourian, and Chi-Te Liang for insightful conversations, and Hava R. Schwartz for developments in our fabrication techniques. We are grateful for the contributions to instrumentation and measurement software by Andrew J. Bestwick, Eli J. Fox, Aaron L. Sharpe, and Menyoung Lee. Device fabrication and measurement was supported by the U.S. Department of Energy, Office of Science, Basic Energy Sciences, Materials Sciences and Engineering Division, under Contract DE-AC02-76SF00515. Infrastructure and cryostat support were funded in part by the Gordon and Betty Moore Foundation through Grant No. GBMF3429. Part of this work was performed at the Stanford Nano Shared Facilities (SNSF), supported by the National Science Foundation under award ECCS-1542152.  The theoretical work in Singapore was funded by the National University of Singapore Young Investigator Award (R-607-000-094-133), and the Singapore Ministry of Education (MOE2017-T2-1-130). The work at Rutgers was supported by the Gordon and Betty Moore Foundation'€™s EPiQS Initiative (GBMF4418) and the National Science Foundation (NSF) (EFMA-1542798). 

\end{acknowledgments}

\end{document}


\title{Supplemental materials for Absence of strong localization at low conductivity in the topological surface state of low disorder \ce{Sb2Te3}}

\author{Ilan T. Rosen}
\affiliation{Department of Applied Physics, Stanford University, Stanford, California 94305, USA}
\affiliation{Stanford Institute for Materials and Energy Sciences, SLAC National Accelerator Laboratory, Menlo Park, California 94025, USA}
\author{Indra Yudhistira}
\affiliation{Department of Physics, National University of Singapore, Singapore 117551, Singapore}
\affiliation{Centre for Advanced 2D Materials and Graphene Research Centre, National University of Singapore, Singapore 117546, Singapore}
\author{Girish Sharma}
\affiliation{Department of Physics, National University of Singapore, Singapore 117551, Singapore}
\affiliation{Centre for Advanced 2D Materials and Graphene Research Centre, National University of Singapore, Singapore 117546, Singapore}
\author{Maryam Salehi}
\affiliation{Department of Materials Science and Engineering, Rutgers, the State University of New Jersey, Piscataway, New Jersey 08854, USA}
\author{M.~A. Kastner}
\affiliation{Department of Physics, Stanford University, Stanford, California 94305, USA}
\affiliation{Stanford Institute for Materials and Energy Sciences, SLAC National Accelerator Laboratory, Menlo Park, California 94025, USA}
\affiliation{Department of Physics, MIT, Cambridge, Massachusetts 02139, USA}
\affiliation{Science Philanthropy Alliance, 480 S. California Ave, Palo Alto, California 94306, USA}
\author{Seongshik Oh}
\affiliation{Department of Physics and Astronomy, Rutgers, the State University of New Jersey, Piscataway, New Jersey 08854, USA}
\author{Shaffique Adam}
\affiliation{Department of Physics, National University of Singapore, Singapore 117551, Singapore}
\affiliation{Centre for Advanced 2D Materials and Graphene Research Centre, National University of Singapore, Singapore 117546, Singapore}
\affiliation{Yale-NUS College, Singapore 138614, Singapore}
\author{David Goldhaber-Gordon}
\affiliation{Department of Physics, Stanford University, Stanford, California 94305, USA}
\affiliation{Stanford Institute for Materials and Energy Sciences, SLAC National Accelerator Laboratory, Menlo Park, California 94025, USA}

\maketitle

\renewcommand{\thetable}{S\arabic{table}}
\renewcommand{\thefigure}{S\arabic{figure}}
\renewcommand{\thesection}{S\arabic{section}}
\renewcommand{\thesubsection}{S\arabic{subsection}}
\renewcommand{\theequation}{S\arabic{equation}}

\setcounter{secnumdepth}{3}

\twocolumngrid
\setcounter{equation}{0}
\setcounter{figure}{0}

\section{Methods}
\begin{figure}[ht]
	\includegraphics[width=0.32\textwidth]{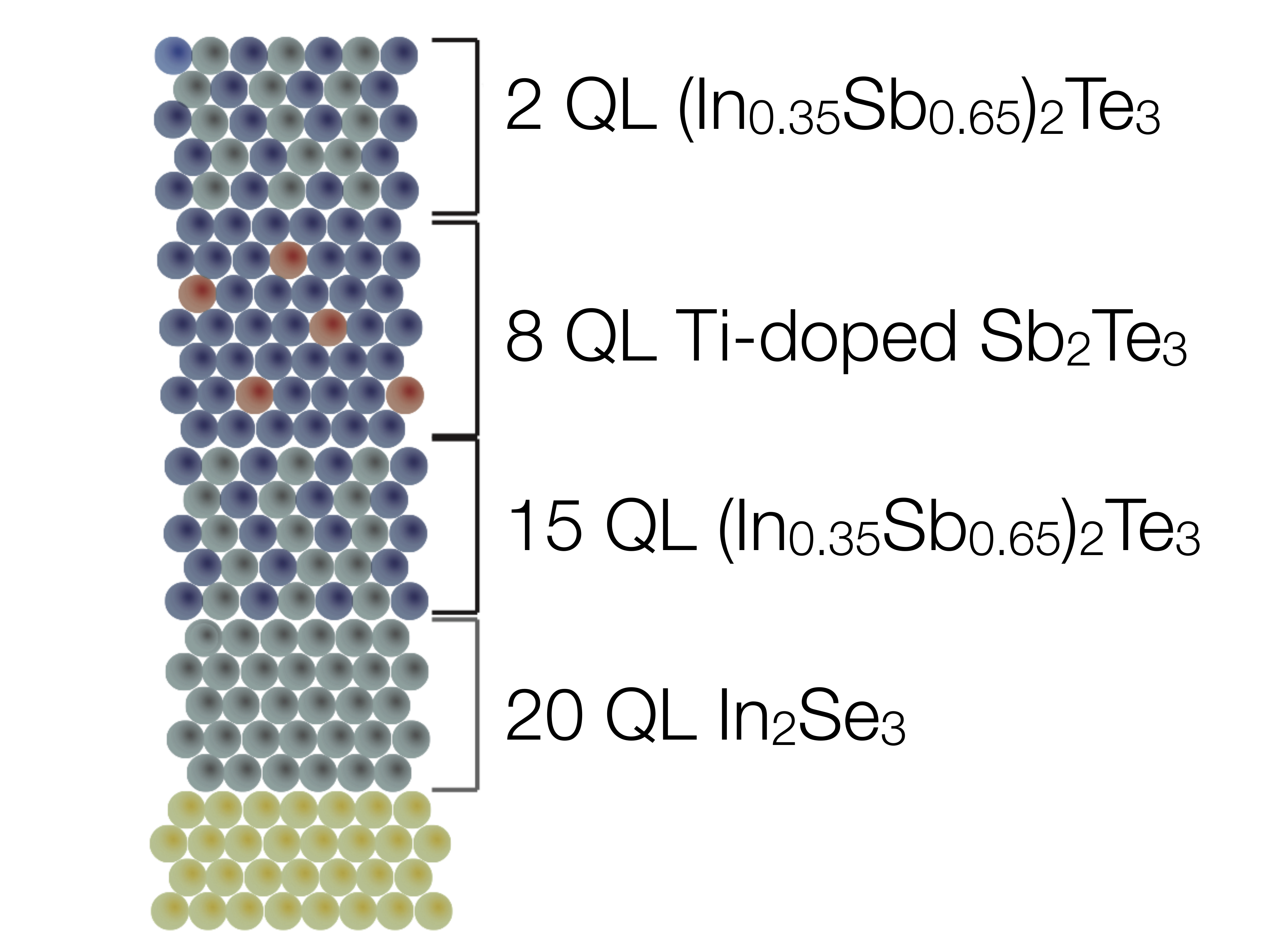}
	\caption{Growth of a low disorder, low carrier density topological insulator. A virtual substrate, consisting of the trivial insulators \ce{In2Se3)} and \ce{Sb_{0.65}In_{0.35}Te3} provides an epitaxially matched template for the \ce{Sb2Te3} topological insulator, lowering disorder. \ce{Ti} counter-doping finely tunes the Fermi level.}
	\label{Fig:Growth}
\end{figure}

\emph{Growth}: Thin-films were grown on sapphire (Al2O3) (0001) substrates by molecular beam epitaxy under ultra-high vacuum. First, a 3 quintuple layer (QL, 1 QL $\approx$ 1 nm) layer of \ce{Bi2Se3} was deposited at  170$^\circ$C, serving as a template for the following 20 QL \ce{In2Se3} layer, deposited at 300$^\circ$C. The \ce{Bi2Se3}/\ce{In2Se3} heterostructure was then heated to 600$^\circ$C, causing the \ce{Bi2Se3} layer to evaporate and diffuse out of the \ce{In2Se3} layer. The remaining \ce{In2Se3} insulating layer served as a template for the next layers. A 15 QL-thick \ce{Sb_{0.65}In_{0.35}Te3} layer was then deposited at 260$^\circ$C, forming a epitaxially matched virtual substrate for the 8 QL \ce{Sb2Te3} layer. After deposition of the \ce{Sb2Te3}, another 2 QL \ce{Sb_{0.65}In_{0.35}Te3} layer was deposited \emph{in situ} as a capping layer at 260$^\circ$C. The thickness of the capping layer was chosen as a compromise to protect the topological layer (thicker cap) while still allowing high gate tunability (thinner cap). The \ce{Sb2Te3} layer was doped with 2\% Ti to finely tune the chemical potential. The resulting material platform is sketched in Fig.~\ref{Fig:Growth}; further details regarding growth may be found in ref.~\cite{Salehi2018}.

\begin{figure}[ht]
	\includegraphics[width=0.3\textwidth]{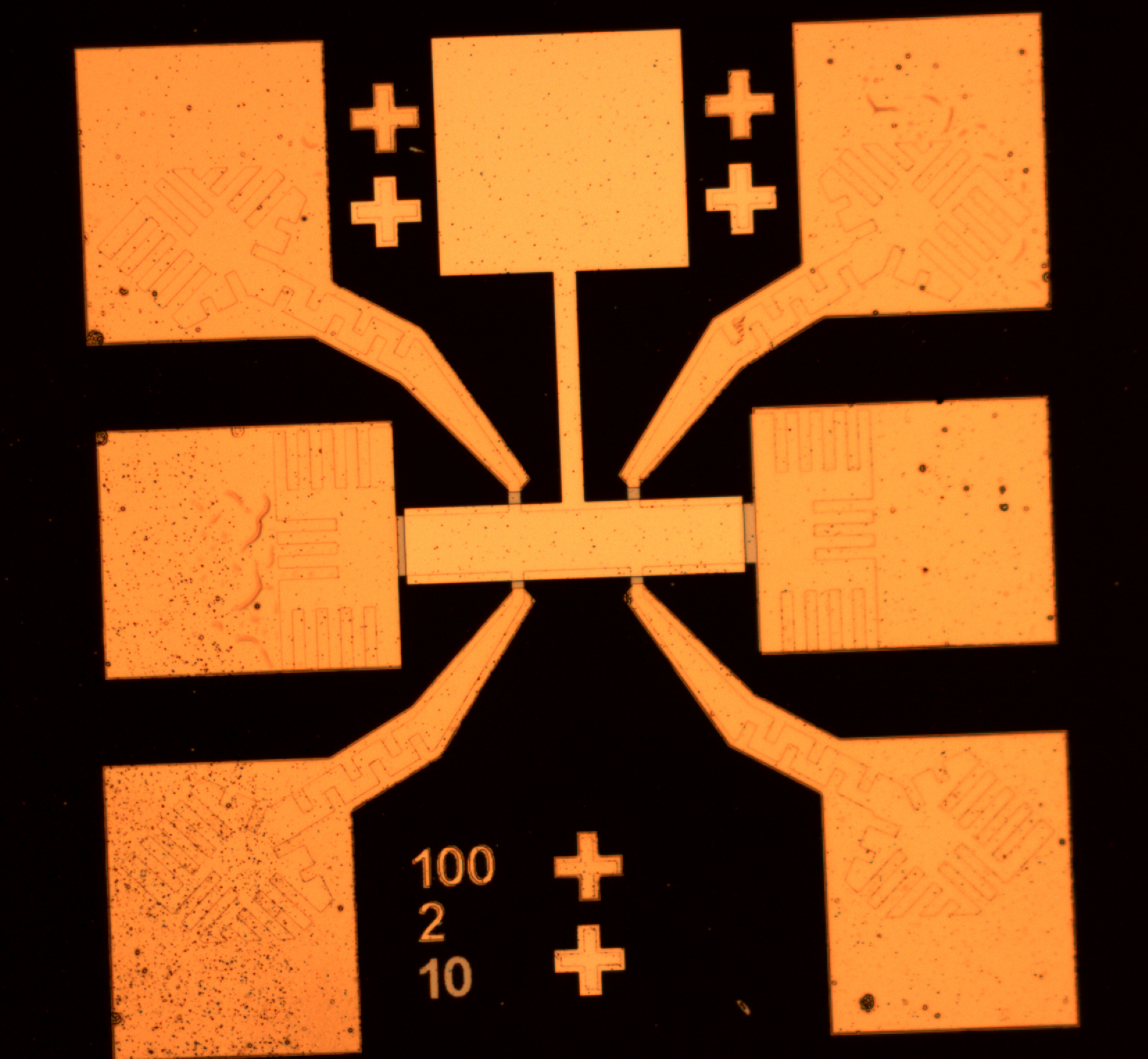}
	\caption{An optical micrograph of the device studied in this work. The device is a 50 $\mu$m wide Hall bar with 100 $\mu$m between voltage terminals. The voltage terminals are 10 $\mu$m wide. The topological insulator is gated through a 40 nm alumina dielectric.}
	\label{Fig:Device}
\end{figure}

\emph{Device fabrication}: The Hall bar device, shown in Fig.~\ref{Fig:Device}, is 50 $\mu\text{m}$ wide with two squares (100 $\mu$m) between the 10 $\mu$m wide voltage terminal leads. It was fabricated using photolithographic patterning. For each patterning step, a hexamethyldisilazane adhesion layer was spin coated, followed by Megaposit SPR 3612 photoresist and a pre-exposure bake at 80$^\circ$C for 180 s, chosen to avoid thermal damage to the film. The photoresist was exposed at approximately 50 mJ/cm$^2$ under an ultraviolet mercury vapor lamp through a contact mask; the photoresist was developed in Microposit developer CD-30 for 35 s. After patterning, the device geometry was defined by a dry etch of the surrounding film with an Ar ion mill. After patterning, Ohmic contacts were made by first cleaning the contact area with a brief \emph{in situ} Ar ion etch, and then evaporating 5 nm Ti and 100 nm Au, followed by liftoff. To realize a robust top gate, a dielectric was grown uniformly across the film by first evaporating a 1 nm Al seed layer, which was allowed to oxidize, and then depositing approximately 40 nm of alumina by atomic layer deposition. The top gate was then patterned and was deposited by evaporating 5 nm Ti and 85 nm Au, followed by liftoff. Excess alumina dielectric on the surrounding area was etched using Microposit developer CD-26 (tetramethylammonium hydroxide based, metal ion free). All metal evaporation was done in a Kurt Lesker electron beam evaporator with an \emph{in situ} Ar ion source. Atomic layer deposition used trimethylaluminum precursor and water as the oxidizer in a nitrogen purged vacuum chamber.

\emph{Transport}: The electrostatic top gate exhibited pronounced hysteresis, characteristic of charge trapping in oxide dielectrics. In particular, upon changing the gate voltage, the system required hundreds of seconds to re-equilibrate, which we attribute to slow tunneling of charge carriers into and out of traps in the oxide. This is discussed later in detail. To account for the dynamics of the gate oxide, gate voltages were always swept upwards starting from $V_g = -10\text{ V}$ and the system was allowed to equilibrate before measurement at each gate voltage.

Measurements were performed in a liquid He-4 cryogenic system ($1.6\text{ K}\leq T \leq 30\text{ K}$ as well as in a He-3/He-4 dilution refrigerator ($30\text{ mK}\leq T\leq 1.2\text{ K}$). Each cryostat was equipped with a 14~T solenoid, with the field direction perpendicular to the plane of the Hall bar. We mainly used standard lock-in techniques for electrical measurements. However, because the resistances observed near the CNP were so large at the lowest temperatures, we measured in DC rather than AC at the lowest temperatures. Here, using a high output impedance current source, the sample was biased with a DC current whose polarity periodically switched (generally, every 2.337 s) and the four-terminal voltage was measured using a high-precision nanovoltmeter; the resistance was taken to be the anti-symmetric component under positive and negative bias polarity. Anti-symmetrization in bias polarity removes spurious thermoelectric contributions from the wiring. All reported resistances are four-terminal voltage measurements divided by the constant current bias. Resistance measurements were made at both positive and negative magnetic fields; all magnetoconductance data is obtained from the symmetrized longitudinal resistance $R_\text{sym}(B) = (R(B) + R(-B))/2$.

The sheet carrier density was calculated from $\rho_{xy}$ by fitting the single carrier model $n=B(e\rho_{xy})^{-1}$ where $e$ is the electronic charge. The carrier mobility $\mu$ was then obtained from the zero-field sheet resistivity as $\mu=(e\rho_{xx}n)^{-1}$. Of the forty-one gate voltage values shown in Fig.~1, the standard error of the fitted Hall slope of three gate voltage values (all near $V_\text{min}$) is greater than the Hall slope itself, a consequence of mixing between the Hall voltage and the much larger longitudinal voltage,
so the sign of the density is not determined. For these cases, extracted density and mobility are not plotted in Fig.~1. In addition, we caution that the extracted mobility, as determined by a single carrier model, is inaccurate near charge neutrality where there are puddles of electrons and holes.

For a Dirac cone, density fluctuations give rise to an energy broadening given by $E = \hbar v_F \sqrt{\pi n}$. The energy scale corresponding to the root mean square (rms) of the density when the average density is $n=0$ is $E_\text{rms}=\hbar v_F \sqrt{\pi n^*}$, with $n^*\equiv \sqrt{n_\text{eff}^e n_\text{eff}^h}=n_\text{rms}/\sqrt{3}$ where $n_\text{eff}^{e(h)}$ is the typical density inside electron (hole) puddles such that the minimum conductivity is $\sigma_\text{min}=n^* e \mu$. The factor of $\sqrt{3}$ comes from an effective mean theory calculation~\cite{Adam2009,Ping2014}.

\section{The Ioffe-Regel violating regime}~\label{Section IR violation}
Neglecting quantum coherent corrections, the conductivity of a metal is known to decrease with rise in temperature. This behavior can be well understood within the semi-classical theory, which suggests that the mean free path of electrons between subsequent collisions reduces as the number of thermal collisions increase with temperature. However, such a reduction in the conductivity of metals eventually saturates~\cite{Ioffe1960}, as mean free path can never
become shorter than the inter-atomic lattice spacing. This is known as the Ioffe-Regel (IR) limit, or sometimes the Mott-Ioffe-Regel (MIR) limit~\cite{Mott1972}, which places a lower bound on the minimum conductivity of metals. Evidence that some high-temperature superconducting compounds violate the IR limit has attracted recent attention~\cite{Hussey2004}. In this section we will attempt to answer the following question: can we theoretically ever obtain violation of the IR limit in a disordered metal? Since our experiment clearly demonstrates a strong breakdown of the IR limit ($\sigma\sim 0.01 e^2/h$), we would like to understand what information this conveys about the material. 

The Boltzmann transport conductivity for massless Dirac fermions has previously been calculated as~\cite{Adam2007}
\begin{align}
\sigma = \frac{e^2}{h}\frac{n^*}{n_{\text{imp}}}\frac{2}{G[2r_s]},
\label{Eq:min_cond}
\end{align}
where the following self-consistency condition needs to be satisfied in order to determine the residual density~\cite{Adam2007}
\begin{align}
\frac{n^*}{n_{\text{imp}}} = 2 r_s^2 C_0^{\text{RPA}}(r_s,4d\sqrt{\pi n^*}),
\label{Eq:selfconsistent1}
\end{align}
and the function 
\begin{align}
&C_0^{\text{RPA}}(r_s,a) = -1 + \frac{4E_1(a)}{(2+\pi r_s)^2} + \frac{2e^{-a}r_s}{1+2r_s} \nonumber\\
&+ (1+2r_s a) e^{2r_sa} (E_1(2r_sa) - E_1(a(1+2r_s))),
\end{align}
is related to the random voltage fluctuations. In the above $E_1(a)$ is the exponential integral function. For an arbitrary $r_s$ and $n_{\text{imp}}d^2$, it is not analytically tractable to obtain the minimum conductivity $\sigma$ due to the complicated form of the function $C_0^{\text{RPA}}$. In principle, this can always be evaluated numerically. Before proceeding with the numerical evaluation, we will first consider some special limiting cases. We note that the evaluation of the function $C_0^{\text{RPA}}$ is in general non-convergent as a series expansion unless we impose a prior assumption on the desired solution $a$. However, such an assumption on $a$ needs to be self-consistent with the actual solution $a$.

In the limit of $r_s\ll 1$, we have 
\begin{align}
C_0^{\text{RPA}}(r_s,a) &\approx -1 + {E_1(a)}\nonumber\\& + e^{2r_sa} (E_1(2r_sa) - E_1(a(1+2r_s))),
\end{align}
We can Taylor expand $E_1(a+2r_sa)$ around $a$ as 
\begin{align}
E_1(a+2r_sa) = E_1(a) - \frac{e^{-a} (2r_s a)}{a} + \mathcal{O}(r_s^2a^2)
\end{align}
This expansion gives us a good approximation when $a\ll 1$. If we assume that $a\ll 1$, then 
the function $C_0$ up to first order in $r_s$  becomes 
\begin{align}
C_0^{\text{RPA}}(r_s,a) &= -1 + E_1(2r_sa) - 2r_s e^{-a} \nonumber\\& + 2r_sa(E_1(2r_sa) -E_1(a))
\end{align}
We will shortly see whether the assumption of $a\ll 1$ gives us a self-consistent solution for $a$ or not.
We are interested in the limit when $r_s$ is very small, therefore we will just retain the dominant terms in $C_0^{\text{RPA}}$. Doing so, we arrive at $
\lim_{r_s\rightarrow 0} C_0^{\text{RPA}}(r_s,a) \approx -\ln(2r_s a) $.
The self consistency condition (Eq.~\ref{Eq:selfconsistent1}) becomes $
-\ln(2r_sa) = {a^2}/{32 \pi n_{\text{imp}} d^2r_s^2}$.
In the limit of $32 \pi n_{\text{imp}} d^2\gg r_s^{-2}$, i.e. high impurity concentration, we can evaluate $a \approx (\sqrt{x+r_s^2}-r_s)/x$, where $x=1/32\pi n_{\text{imp}}d^2r_s^2$. For our solution to be consistent with our prior assumption of $a\ll1$ we must have $x+2r_s\gg 1$. By our definition of the high impurity concentration limit $x\ll 1$. As we have noted we are also interested in the limit $r_s\ll 1$. We have arrived at a contradiction: for high impurity concentration and $r_s\ll 1$, we see by inspection that there does not exist a self-consistent solution $a$, such that $a\ll1$. To determine the existence of a self-consistent solution in the low impurity concentration limit is not analytically feasible. Further, since we are primarily interested in the IR violating regime, which is physically expected only when there is large disorder, we do not pursue this case here.  
To evaluate the limit of $a\gg 1$  we start with the complete screening approximation ($r_s a\gg 1$), where the function $ C_0^{\text{RPA}}(r_s,a) = {1}/{4 a^2 r_s^2}$~\cite{Adam2007}. For $r_s\ll 1$, this implies $a\gg r_s^{-1}\gg 1$. The self-consistency condition gives us $a^4 = 8\pi n_{\text{imp}}d^2$. For this solution to be consistent with our prior assumption on $a$, we must have $8\pi n_{\text{imp}}d^2\gg r_s^{-4}\gg 1$. Therefore for $r_s\ll 1$ it is always possible to find a self-consistent solution $a$ such that $a\gg r_s^{-1}\gg 1$ as long as the impurity concentration is large enough. We can now evaluate the minimum conductivity for this case, which is given by 
\begin{align}
\sigma = \frac{1}{\sqrt{8\pi n_{\text{imp}}d^2 r_s^4}}\frac{e^2}{h} \ll \frac{e^2}{h}
\end{align}
It is straightforward to conclude that when $8\pi n_{\text{imp}}d^2\gg r_s^{-4}\gg 1$, we are guaranteed to find a solution $a$ such that $a\gg r_s^{-1}\gg 1$, and for such a solution $\sigma\ll e^2/h$, i.e., it violates the Ioffe-Regel limit. 
\begin{figure}[ht]
	\includegraphics[scale=0.12]{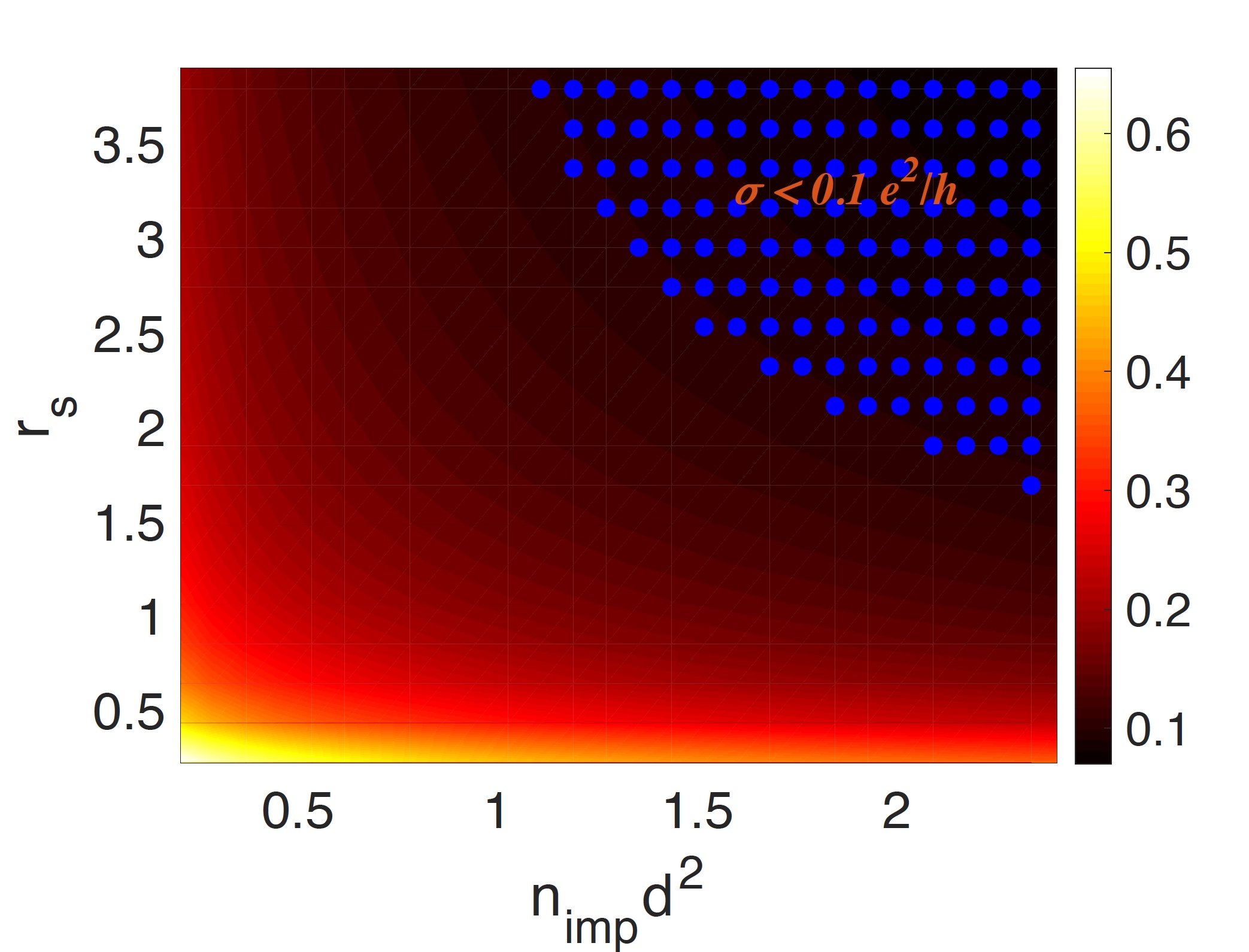}
	\includegraphics[scale=0.12]{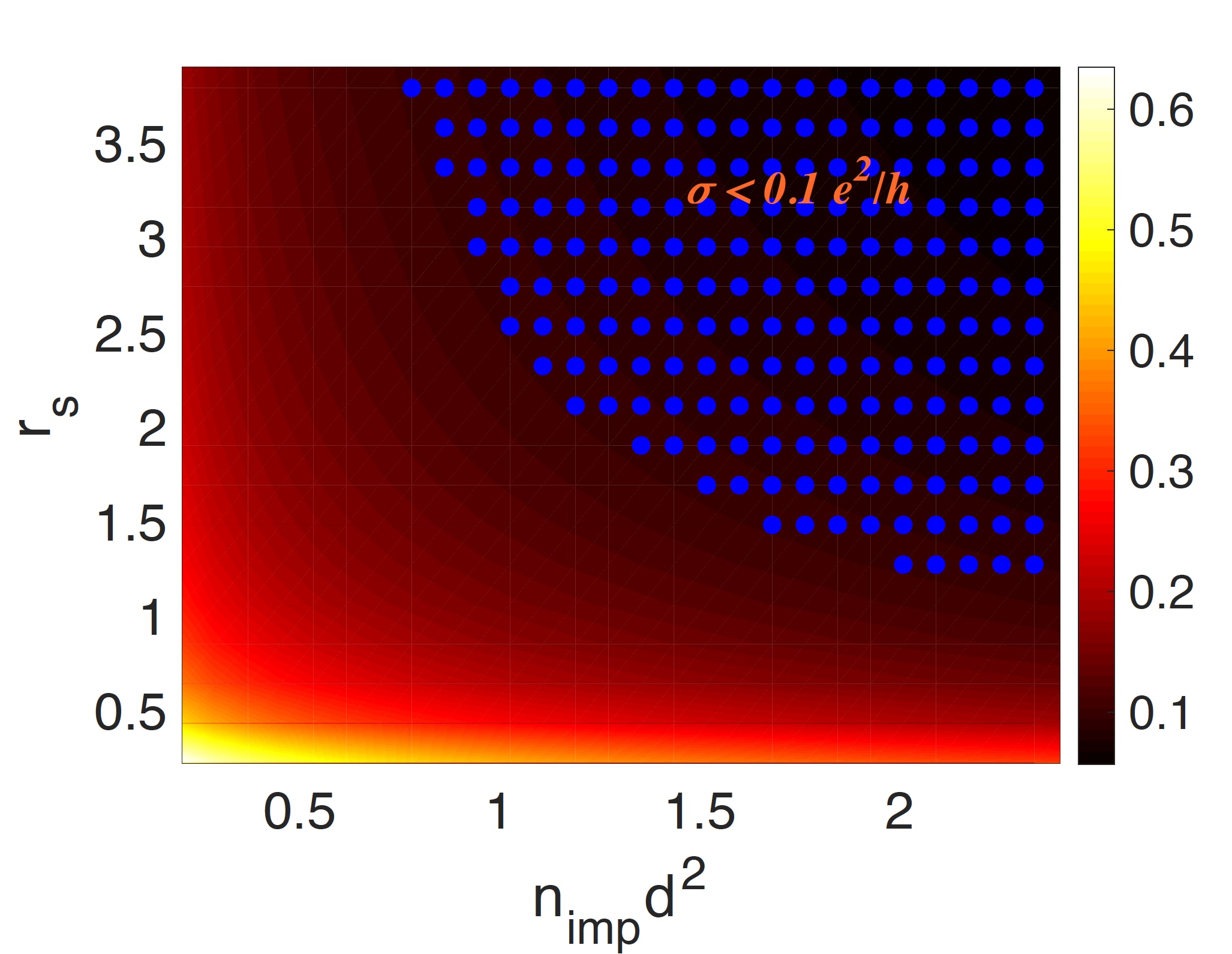}
	\caption{Numerically obtained minimum conductivity as a function of $r_s$ and $n_{\text{imp}}d^2$ by solving Eq.~\ref{Eq:Suppl_EMT},~\ref{Eq:Suppl_sigmaB} and ~\ref{Eq:Suppl_SelfCons}. The top and the bottom panels are for the cases $n_0= 0$ (symmetric) and $n_0\neq 0$ (asymmetric, $n_0\sim 400 \times 10^{10} $cm$^{-2}$), respectively. The dotted blue region gives us an estimate of the value of $r_s$ and $n_{\text{imp}}d^2$ for which $\sigma < 0.1 e^2/h$. Note that for large $r_s$ and $n_{\text{imp}}d^2$  we are in the strong IR violating regime ($\sigma\ll e^2/h$), as noted in Table~\ref{table:nonlin}. The presence of an asymmetry parameter $n_0$ reduces the threshold value for $n_{imp}d^2$ for which the violation of the Ioffe-Regel limit can be observed.}
	\label{Fig:Phase_IR}
\end{figure}

\begin{table}[ht]
	\centering 
	\begin{tabular}{c c c c c} 
		\hline\hline 
		$r_s$ & $a$ & s-c & $\text{IR violation}$ & condition\\ [0.5ex] 
		\hline\hline 
		$\gg 1$ & $\gg 1$ & Yes & Yes & $n_{\text{imp}} d^2\gg 1$\\ 
		$\ll 1$ & $\gg 1$ & Yes & Yes & $n_{\text{imp}} d^2 r_s^4\gg 1$\\
		$\gg 1$ & $\ll 1$ & No  & - & -\\ 
		$\ll 1$ & $\ll 1$ & No  & - & -\\
		[1ex] 
		\hline 
	\end{tabular}
	\caption{Conditions which guarantee the violation of the IR limit. The third, fourth, and fifth column represent: if the self-consistent (s-c) solution is possible or not with our assumption on $a$, whether there is violation of IR limit (if there is a s-c solution), and the corresponding condition for violation (if there is one), respectively. } 
	\label{table:nonlin} 
\end{table}

Let us now discuss the case when $r_s\gg 1$. Again, if we first assume that $a\gg 1$, then we directly end up in the complete screening limit ($r_s a\gg 1$), where $ C_0^{\text{RPA}}(r_s,a) = {1}/{4 a^2 r_s^2}$~\cite{Adam2007}. The self-consistency condition gives us $a^4 = 8\pi n_{\text{imp}}d^2$. For this solution to be consistent with our prior assumption on $a$, we must have $8\pi n_{\text{imp}}d^2\gg 1$. The minimum conductivity in this case is given by 
\begin{align}
\sigma \leq \frac{10}{\sqrt{32 \pi n_{\text{imp}}d^2}}\frac{e^2}{h}\ll \frac{e^2}{h}
\end{align}
where we have used a bound on $G(x)$ for large $x$: $ G(x)\leq 0.2$.
Again, we can conclude that when $n_{\text{imp}}d^2\gg 1$ and $r_s\gg 1$, we are guaranteed to find a self consistent solution $a$ such that $a\gg 1$, and for such a solution $\sigma\ll e^2/h$, i.e., it violates the Ioffe-Regel limit. 
When $r_s\gg 1$, and if we now assume that $a\ll 1$ we have $ C_0^{\text{RPA}}(r_s,a) = -{4\ln(\gamma a)}/{\pi^2r_s^2}$, $\gamma$ being the Euler's constant. The self consistency condition becomes $\gamma a = \exp(-a^2\pi/128n_{\text{imp}}d^2)$. When $128n_{\text{imp}}d^2\gg 1$, i.e., for high impurity concentration, we obtain $a = (\sqrt{\gamma^2 + 4y}-\gamma)/2y$, where $y=\pi/128n_{\text{imp}}d^2$. For our solution to be consistent with the prior assumption on $a\ll 1$, we must have $1\ll y+\gamma$. Since $y>0$ and $y\ll 1$, the condition $1\ll y+\gamma$ is never satisfied because 1 and $\gamma$ are on the same order of magnitude. We can thus conclude that in the limit $r_s\gg 1$, we cannot find a self consistent solution $a$, such that $a\ll 1$, at least in the high impurity concentration limit. 

We can now generalize our results to obtain the conditions for violation of IR limit, which are summarized in Table~\ref{table:nonlin}. 
In Fig.~\ref{Fig:Phase_IR} we plot the numerically obtained minimum conductivity (using Eq.~\ref{Eq:Suppl_EMT}, ~\ref{Eq:Suppl_sigmaB} and ~\ref{Eq:Suppl_SelfCons}) as a function of $r_s$ and $n_{\text{imp}}d^2$ for both $n_0=0$ and $n_0\neq 0$, $n_0$ being the asymmetry parameter between the electron and hole bands~\cite{Adam2012}. We note that for large $r_s$ and $n_{\text{imp}}d^2$ we are in the strong IR violating regime ($\sigma\ll e^2/h$), which validates the conclusion in Table~\ref{table:nonlin}. In this discussion, the dimensionless parameter $n_{\text{imp}}d^2$ should be thought of in the thermodynamic limit, where $n_{imp}\rightarrow \infty$ as $d\rightarrow 0$ such that $n_{\text{imp}}d^2$ remains constant at its specified value. We also note that the presence of an asymmetry parameter $n_0$ reduces the threshold value for $n_{imp}d^2$ for which the violation of the Ioffe-Regel limit can be observed.


\section{Estimation of $n_\mathrm{rms}$ from experiment\label{sec:nrmsfit}}

Charge density fluctuations ($n_\mathrm{nrms}$) are determined from the fit of Hall coefficient ($R_H$) vs carrier density ($n$) at low temperature. Theoretical $R_H$ is calculated from $R_H^\mathrm{EMT}=\lim_{B\rightarrow 0} \rho_{xy}^\mathrm{EMT}/(e B)$, where the effective medium theory (EMT) conductivity is obtained from solving the following coupled equations
\begin{align}
\int dnP\frac{\sigma_{xx}^{2}[n]-\left(\sigma_{xx}^{\mathrm{EMT}}\right)^{2}+\left(\sigma_{xy}^{\mathrm{EMT}}-\sigma_{xy}[n]\right)^{2}}{\left(\sigma_{xx}^{\mathrm{EMT}}+\sigma_{xx}[n]\right)^{2}+\left(\sigma_{xy}^{\mathrm{EMT}}-\sigma_{xy}[n]\right)^{2}} & =0\nonumber\\
\int dnP\frac{\sigma_{xy}[n]-\sigma_{xy}^{\mathrm{EMT}}}{\left(\sigma_{xx}^{\mathrm{EMT}}+\sigma_{xx}[n]\right)^{2}+\left(\sigma_{xy}^{\mathrm{EMT}}-\sigma_{xy}[n]\right)^{2}} & =0,
\label{Eq:Suppl_EMT}
\end{align}
where $P=P\left[n,n_{g},n_{\mathrm{rms}}\right]$ is a Gaussian distribution centered at average carrier density $n_g$ with width given by carrier density fluctuations $n_\mathrm{nrms}$~\cite{Ping2014}.

Note that resistivity matrix is given by the inverse of conductivity matrix, hence
$\rho_{xy}^\mathrm{EMT}=-\sigma_{xy}^\mathrm{EMT}/[\left(\sigma_{xx}^\mathrm{EMT}\right)^2+\left(\sigma_{xy}^\mathrm{EMT}\right)^2]$.

We have used the following longitudinal and transverse conductivity of the single channel model as an input to the EMT equation
\begin{align}
\sigma[n,B]&=\frac{\sigma_B[n]}{1+(\mu_{e(h)}B)^{2}}\begin{pmatrix}1 & \mp\mu_{e(h)}B\\
\pm\mu_{e(h)}B & 1
\end{pmatrix}\\
\sigma_B[n]&=|n|e\mu_{e(h)}\nonumber
\end{align}

$R_H$ reveals $n_\mathrm{rms}$ and ratio of mobility $\mu_e/\mu_h$, rather than each component of mobility separately. From the fit, we infer $n_\mathrm{rms}=(12\pm 2)\times 10^{10}~\mathrm{cm}^{-2}$ and $\mu_e/\mu_h=0.5\pm 0.1$ (see Fig.~1).

\begin{figure}[htbp]
\begin{centering}
\includegraphics[width=.9\columnwidth]{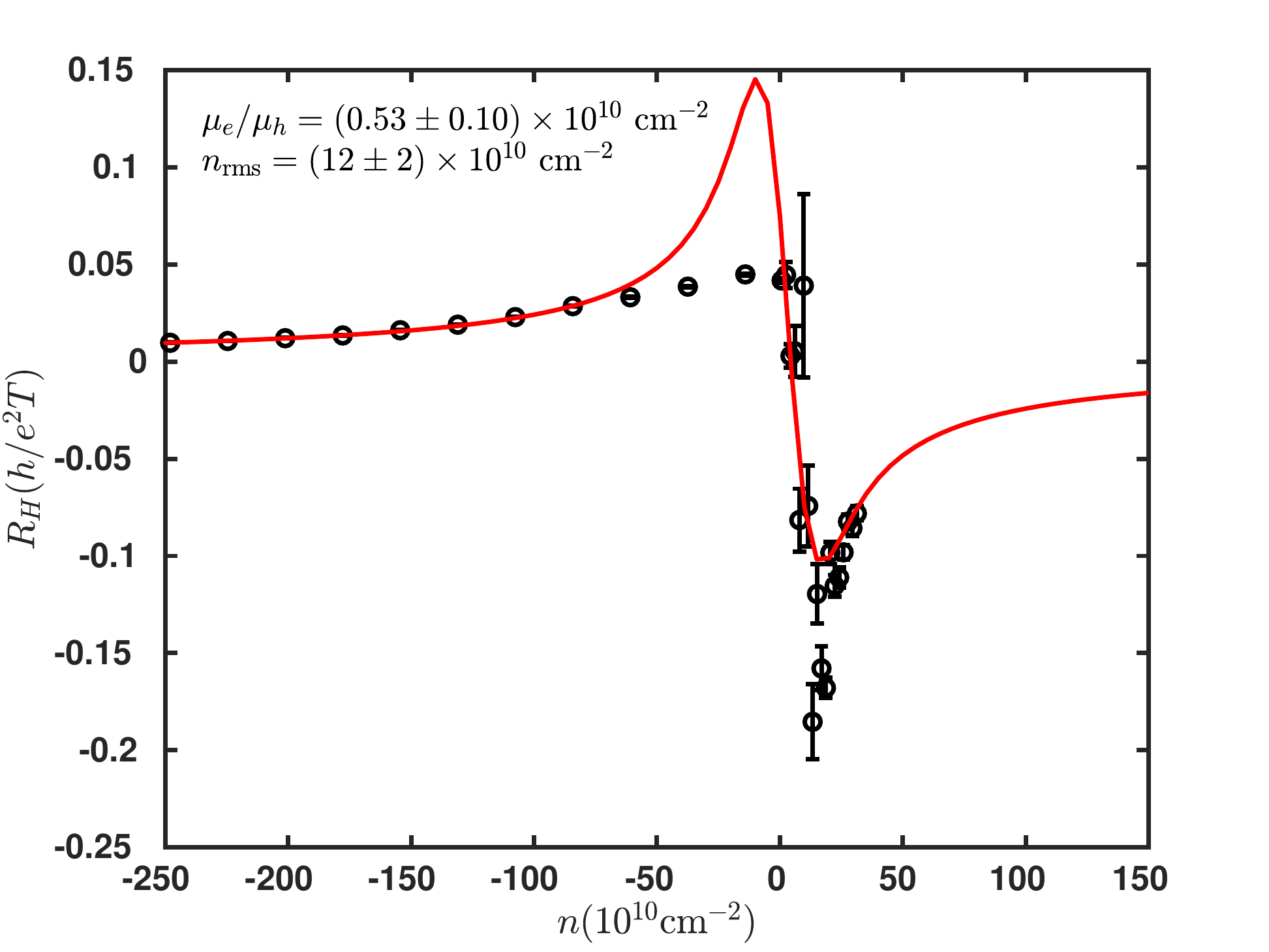}
\par\end{centering}
\caption{Fitting of $R_H$ vs $n$ with $n_\mathrm{rms}$ using minimal model\label{fig:fitRH}}
\end{figure}

\section{Estimation of $r_s$ from experiment}
The effective fine structure constant $r_s$ is estimated by performing simultaneous fitting to both $\sigma_{xx}(B=0)$ and $R_H$, with the value of $n_\mathrm{rms}$ fixed from the fit in section 1. However, unlike section 1 which uses the minimal model, in order to estimate $r_s$, we use a microscopic model of charged impurity scattering with asymmetry parameter $n_0$ as input to the EMT equation~\cite{Adam2012}. This microscopic model is parametrized by 4 parameters: impurity density $n_\mathrm{imp}$, asymmetry parameter $n_0$, distance from the substrate to the sample $d$, and effective fine structure constant $r_s$.
\begin{align}
\sigma_{B}\left[n,n_{0},r_{s}\right] & =\frac{1}{8}\frac{e^{2}}{h}\frac{n}{n_{\mathrm{imp}}}\frac{1}{F_{1}\left[\eta r_{s}/2\right]} \label{Eq:Suppl_sigmaB}\\ \frac{F_{1}[x]}{x^{2}}  =\frac{\pi}{4}+&3x-\frac{3x^{2}\pi}{2}+x\left(3x^{2}-2\right)\frac{\arccos[1/x]}{\sqrt{x^{2}-1}}\nonumber\\
\eta\left[n/n_{0}\right] & =\frac{\sqrt{n_{0}}}{\sqrt{n_{0}}+\operatorname{sgn}(n)\sqrt{|n|}}\nonumber
\end{align} 
Since $n_\mathrm{rms}$ is fixed from the previous fit at section S1, our fitting parameters now are only $n_\mathrm{imp}$, $n_0$, and $r_s$. Distance $d$ is deduced from the self-consistent theory of Ref.~\cite{Adam2012}.
\begin{align}
\frac{y^{2}}{4}+y+sy^{3/2} & =AC_{0}\left[\frac{B\sqrt{y}}{1+s\sqrt{y}}\right]\nonumber\\
C_{0}[x] & =\partial_{x}\left[xe^{x}\int_{x}^{\infty}t^{-1}e^{-t}dt\right]\label{Eq:Suppl_SelfCons}\\
A & =\frac{1}{2}\frac{n_{\text{ imp }}}{n_{0}}r_{s}^{2},\quad B=2r_{s}d\sqrt{4\pi n_{0}},\nonumber
\end{align} 
where $y=n_\mathrm{eff}/n_0$ and $s=\operatorname{sgn}(n)$ denotes the electron (hole) bands for $s=1(-1)$, respectively. We have used the relation $n_\mathrm{rms}=\sqrt{3}n^\star$, where $n^\star=\sqrt{n_\mathrm{eff}^e n_\mathrm{eff}^h}$.

Using this procedure, we obtain $r_s=1.3\pm0.8$, $d=4.7$ nm, $n_\mathrm{imp}=(48\pm 28)\times 10^{10}~\mathrm{cm}^{-2}$, and $n_0=(494\pm 303)\times 10^{10}~\mathrm{cm}^{-2}$ (see Fig.~\ref{fig:fitrs}).

\begin{figure}[htbp]
\begin{centering}
\includegraphics[width=.9\columnwidth]{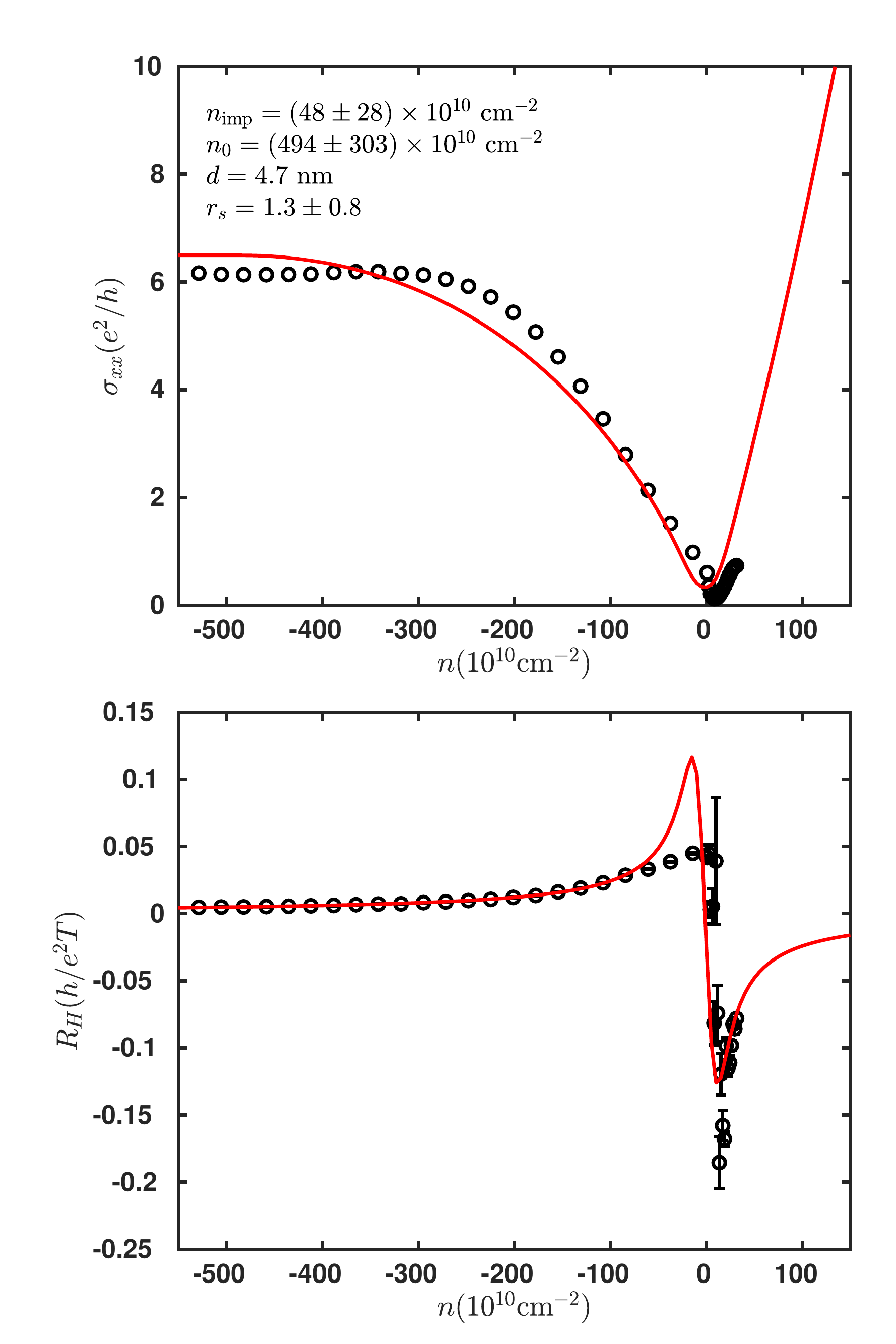}
\par\end{centering}
\caption{Simultaneous fitting of $\sigma_{xx}(B=0)$ vs $n$ and $R_H$ vs $n$ with $n_\mathrm{rms}$ fixed from fit in section \ref{sec:nrmsfit} \label{fig:fitrs}.\label{fig:fitrs}}
\end{figure}

The value of $n_\mathrm{imp}d^2$ that is inferred from the fitting is approximately $0.1\pm 0.06$, not $\gg 1$. Yet the minimum conductivity of our system empirically violates the Ioffe-Regel limit. We point out that the condition $n_{imp} d^2 \gg 1$ guarantees an Ioffe-Regel violating metal in the symmetric case ($n_0=0$), but is not a necessary condition, as seen in Fig.~\ref{Fig:Phase_IR}.
Further, the presence of an asymmetry parameter $n_0$ reduces the threshold value for $n_{imp}d^2$ for which the violation of the Ioffe-Regel limit can be observed. 

\section{Estimation of the surface gap from magnetoconductance}
In a 3D TI, when the hybridization between the top and bottom surfaces is weak, the quantum corrections to the low field magnetoconductance are negative, and electrons exhibit perfect weak antilocalization given by the Hikami-Larkin-Nagaoka (HLN) formula 
\begin{equation}\label{eq:HLN1Cs}
\Delta\sigma = -\frac{e^2}{2\pi h} F\left(\frac{B}{B_\phi}\right),
\end{equation}
where $F(x) = \ln(x) + \Psi(x+\frac{1}{2})$, $\Psi$ being the digamma function, and $B_\phi=\hbar/(4el_\phi^2)$ being the characteristic field associated with the electron coherence length $l_\phi$. The weak antilocalization effect can be ascribed to the $\pi$ Berry phase of the 2D Dirac dispersion, since backscattering effects are 
suppressed by this additional phase factor. This system belongs to the symplectic Wigner-Dyson class (or the AII class) akin to the spin-orbit coupled 2DEG. In contrast to a normal 2DEG (orthogonal Wigner-Dyson class or the AI class), for gapless 2D Dirac fermions the in-plane spin-momentum correspondence requires a spin-flip for backscattering of electrons. The summation of incoming and outgoing spins is 0, leading to the weak antilocalization from the singlet Cooperon mode.

In the limit of a large Dirac mass, spin is polarized along the $z$ axis near the band edges. In this case, the total spin before and after backscattering is 1, which corresponds to one of the triplet Cooperons and leads to weak localization. 
In thin films where the top and bottom surface states hybridize, the Berry phase $\phi_b$ deviates from $\pi$ as 
\begin{align}
\phi_b = \pi \left(1-\frac{\Delta}{2E_F}\right),
\end{align}
where $E_F$ is the Fermi level, due to a finite surface gap (mass term) $\Delta$ arising from surface hybridization. 
The Berry phase induces a crossover from perfect WAL ($\phi_b=\pi$) in the massless (relativistic) limit to perfect WL ($\phi_b=0$) in the large mass (non-relativistic) regime. Localization results in the low field enhancement of conductivity, and the quantum corrections are again given by the HLN formula
\begin{equation}\label{eq:HLN2Cs}
\begin{split}
\Delta\sigma=&-\frac{1}{2\pi}\frac{e^{2}}{h}\left[F\left(\frac{B}{B_{\phi}}\right)-2F\left(\frac{B}{B_{\phi}+B_\mathrm{\Delta}}\right)\right.\\
&\left.-F\left(\frac{B}{B_{\phi}+2B_\mathrm{\Delta}}\right)\right],
\end{split}
\end{equation}
where $B_i=\hbar/(4el_i^2)$ are the characteristic fields associated with the coherence length $l_\phi$, and the crossover length scale $l_\Delta$.  An interesting question arises here: when will one observe a WAL-WL crossover in a TI, or rather what values of $\Delta$ will give rise to WAL-WL crossover in an experiment? This corresponds to the zeros of equation (\ref{eq:HLN2Cs}), which in general are not analytically tractable, but can be obtained numerically.

To address this, let us examine the relevant length scales for carriers in our experimental system. We call the length scale corresponding to the largest magnetic field in our experiment $l_{\text{expt}}$. This is the shortest length scale we probe, and must be much shorter than $l_\phi$ to enable probing coherent transport. Another important scale is the scattering length $l_\Delta$ associated with surface hybridization of the TSS. Any scattering length scale larger than the phase-coherence length cannot be probed in our experiment. Therefore we must satisfy $l_\phi>l_\Delta$ (or equivalently $B_\phi<B_\Delta$) in order to determine the presence of a finite surface gap. Likewise, the effective magnetic field corresponding to a scattering mechanism should also be much smaller than the largest experimental magnetic fields ($B_{\text{expt}}= 14$T here). We thereby have the condition $B_\phi<B_\Delta\ll B_{\text{expt}}$.
 
 To proceed, we will assume $B_\Delta=s B_\phi$, where $s\geq 1$. Around $B=0$, Eq.~\ref{eq:HLN2Cs} can be expanded as 
\begin{align}
\Delta\sigma \approx -\frac{e^{2}}{2\pi h} \frac{x_0^2}{24}\left(1 - \frac{2}{(s+1)^2} - \frac{1}{(2s+1)^2}\right),
\end{align}
where we have defined $x_0=B/B_\phi$. The R.H.S. of the above equation is always negative, implying WAL correction to the conductivity for small magnetic fields (as observed). For large $B$ fields (however small compared to $B_{\text{expt}}$), Eq.~\ref{eq:HLN2Cs} can be expanded as
\begin{align}
\Delta\sigma \approx -\frac{e^{2}}{2\pi h}( 4+\ln((s+1)^2(2s+1)) -2\ln x_0)
\end{align}
Clearly when $B>B_c$ we have a well defined crossover from WAL to WL, where $B_c$ is given by the solution of the following equation:
\begin{align}
2\ln(B_c/B_\phi) \approx 4+\ln((s+1)^2(2s+1)) 
\end{align}
For typical coherence lengths $l_\phi\approx100$~nm, and $s=1$ (the minimum value of $s$ for observing a crossover), this gives us $B_c\approx 0.4$~T. When $s=6$, this gives us $B_c\approx 3\text{ T}\ll B_{\text{expt}}$. If $B_c$ were much larger it would be hard for us to observe the full crossover below our maximum applied field $B_{\text{expt}}=14$T. Hence, to observe a clear signature of WAL-WL crossover in our experiment, we constrain $s$$\in$$(1,6)$. This places a bound on the values of the surface gap $\Delta$ that can be effectively probed since $s\sim (\Delta/v_F)^2$. We find that experiments like ours can exhibit clear signatures of WAL-WL crossover for a surface gap $\Delta$ ranging from $3.3$ to $8$ meV. For smaller values of $\Delta$, the intersurface scattering time becomes greater than the coherence time of the electrons, while for much larger values of the $\Delta$, WAL signatures will be completely washed out. Our estimate in the main text of $\Delta\approx \hbar v_F/l_\Delta = 6.7$~meV falls within this bound.

\section{Extended data}

\begin{figure}[ht]
    \includegraphics[width=0.45\textwidth]{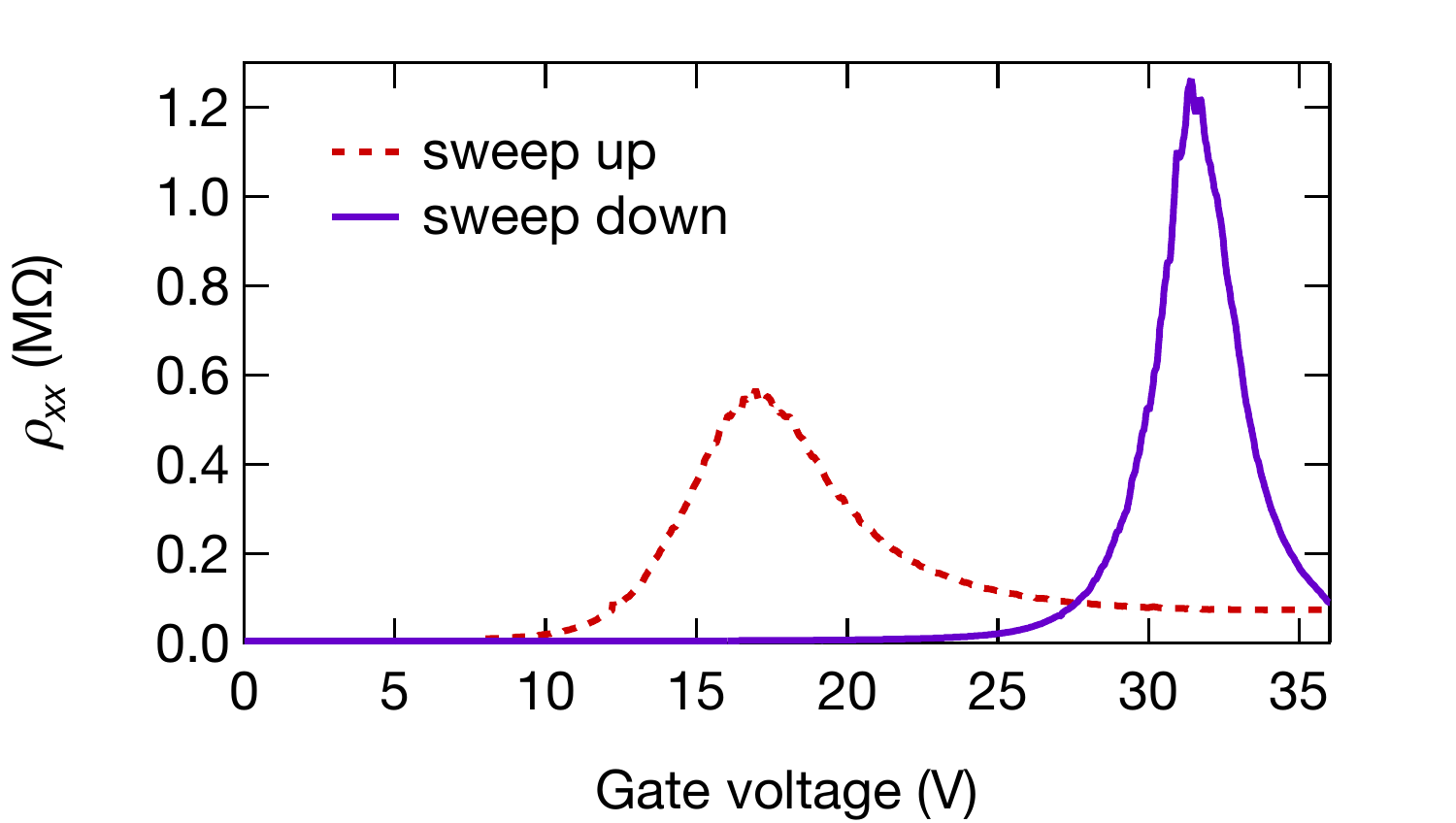}
    \caption{Gate hysteresis. The longitudinal resistivity at $T=29$~mK as the gate voltage is (red) swept up and (purple) swept back down.}
    \label{Fig:hyst}
\end{figure}

\begin{figure}[ht]
    \includegraphics[width=0.48\textwidth]{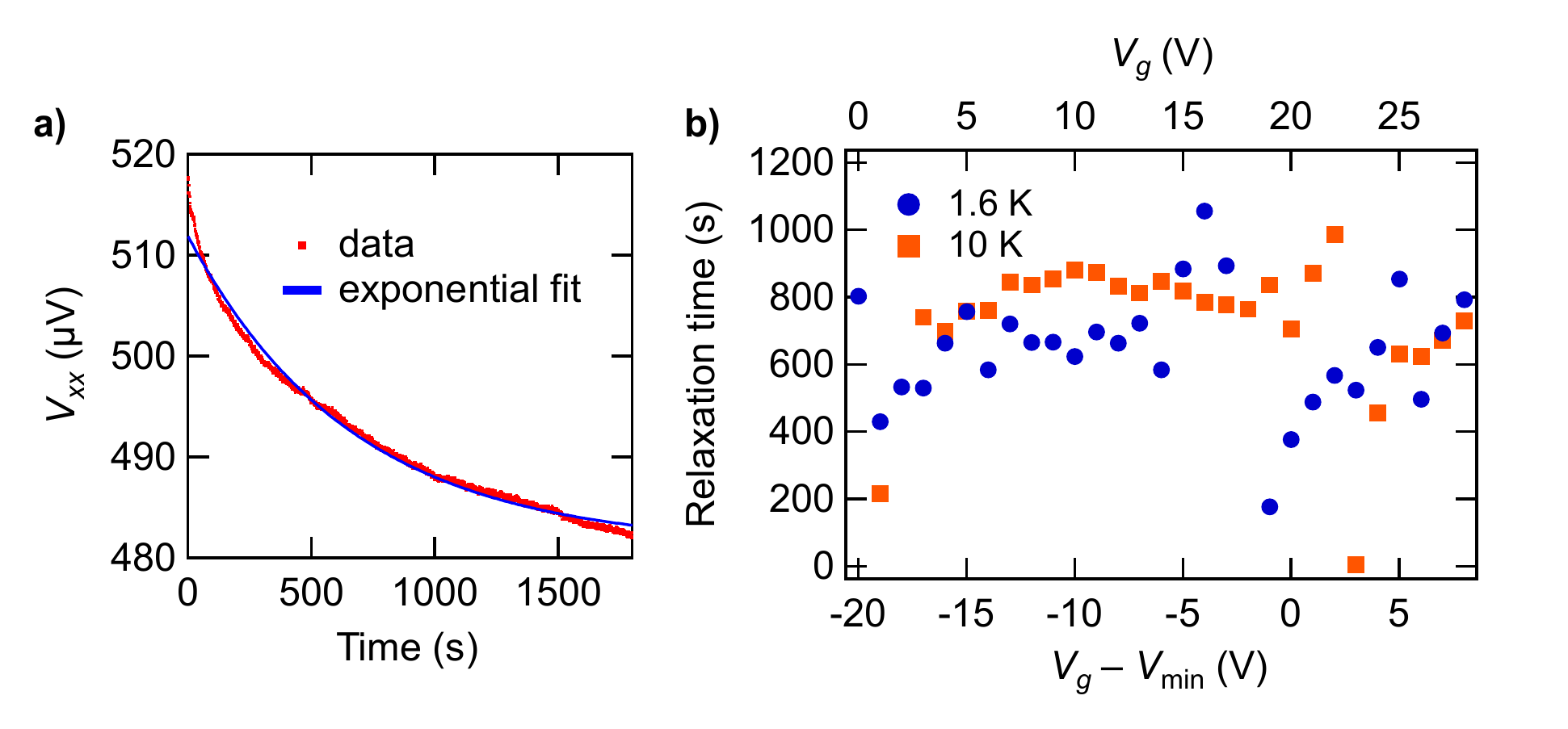}
    \caption{Gate relaxation. a) Upon changing the gate voltage (here, from $V_g=11$~V to 12~V; or from $\Delta V_g = -9$~V to $-8$~V), the longitudinal voltage overshoots its new, higher value and then relaxes. The relaxation fits well to an exponential with a timescale $\tau\sim 1000$~s. b) The fit relaxation timescale $\tau$ versus gate voltage, shown at $T=1.6$~K and 10~K. $\tau$ lacks strong gate voltage dependence.}
    \label{Fig:equil}
\end{figure}

\begin{figure}[ht]
    \includegraphics[width=0.45\textwidth]{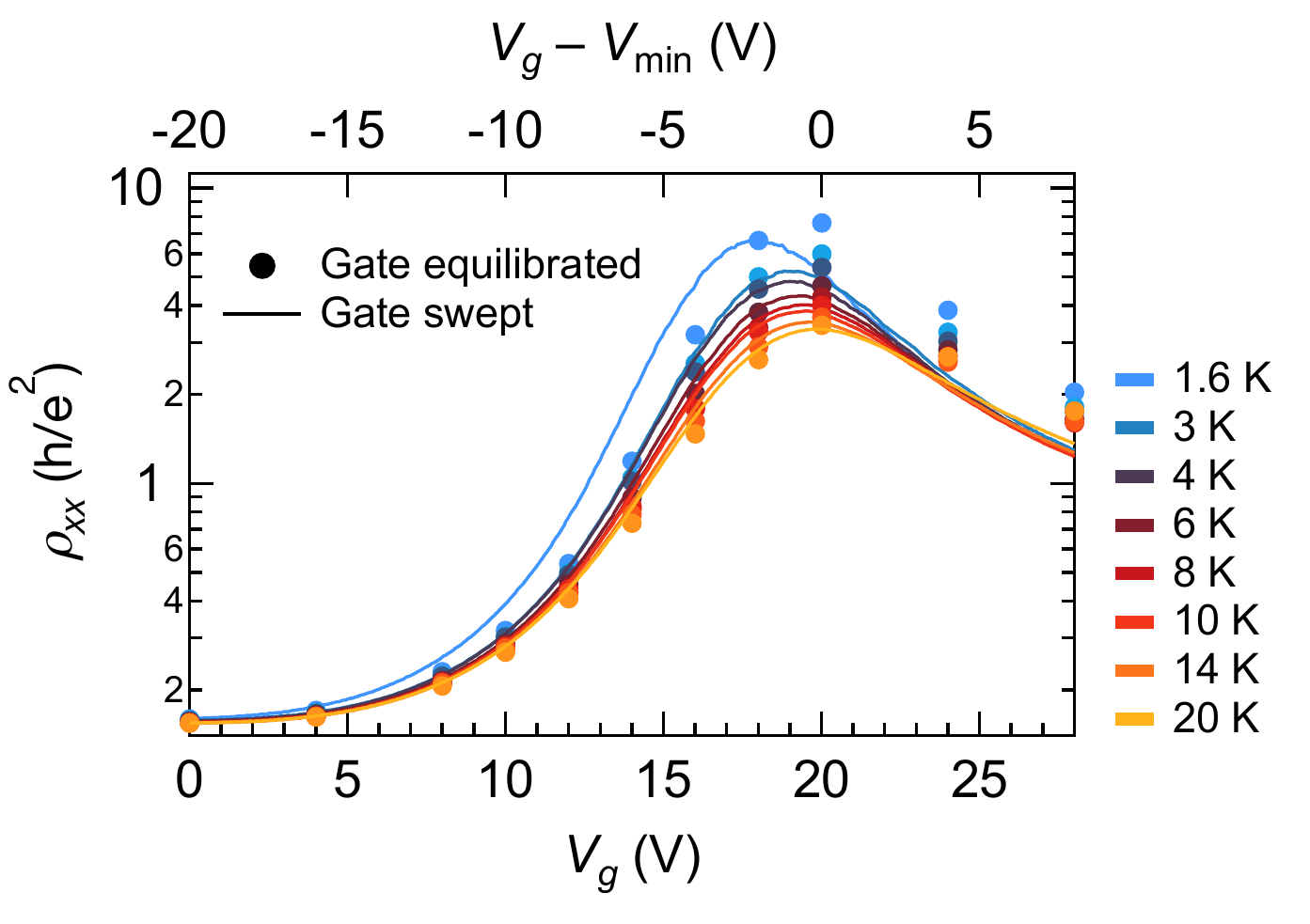}
    \caption{The resistivity versus gate voltage is shown (solid lines) while sweeping the gate and (circles) when the gate oxide is allowed to equilibrate. Data are shown at various temperatures.}
    \label{Fig:sweepvsequil}
\end{figure}

The effect of the top gate exhibited pronounced advanced hysteresis, as shown in Fig.~\ref{Fig:hyst}. We interpret this behavior, which is typical of oxide dielectric gates, as a consequence of electric field screening from charge traps in the gate oxide. To avoid measurement inconsistencies, the gate was always swept from $V_g=-10$~V toward positive gate voltage values. Upon changing the gate voltage, measured resistance initially overshot its equilibrium value at the new gate voltage, and then relaxed exponentially with a time constant of order 1000~s, as described in Fig.~\ref{Fig:equil}. The relaxation rate depends strongly on neither temperature nor gate voltage. This suggests that the relaxation is not associated with RC-type carrier equilibration in the topological insulator (whose resistance depends strongly on temperature and gate voltage), nor is it associated with a thermally activated process in the dielectric. We therefore believe that the relaxation is associated with charges tunneling between charge traps in the gate oxide. To avoid any inconsistencies, all data shown in the main text were taken after allowing transport properties to equilibrate at each new gate voltage. Fig.~\ref{Fig:sweepvsequil} compares the resistivities during a gate sweep and when the gate was allowed to equilibrate.

\begin{figure}[ht]
    \includegraphics[width=0.45\textwidth]{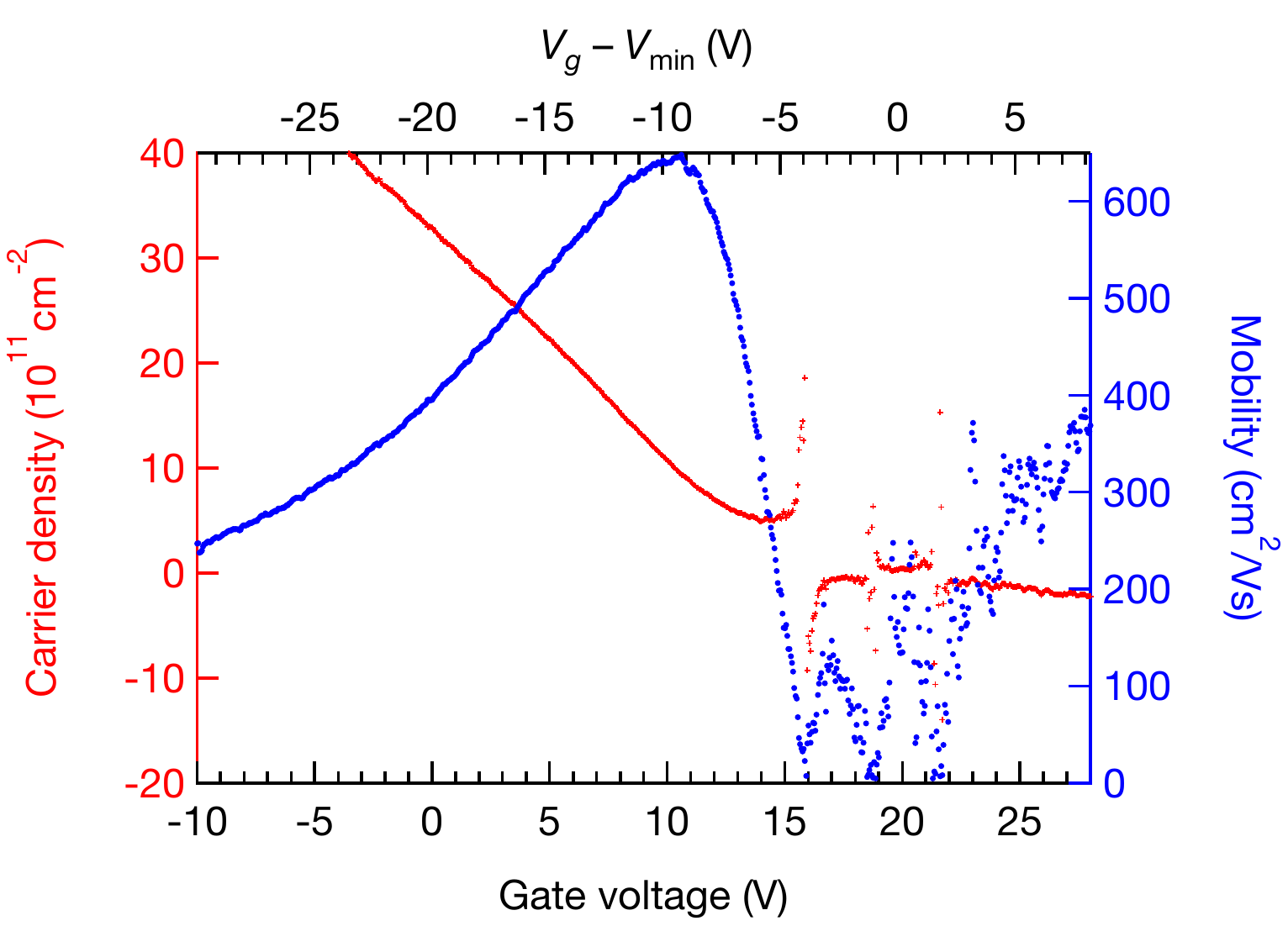}
    \caption{The (red dots, left axis) carrier density $n$ and (blue dots, right axis) mobility are shown as a function of gate voltage at $T=36$~mK as the gate voltage is swept. The density and mobility are extracted from gate sweeps at fields between $-1$ and $1$~T.}
    \label{Fig:nmu}
\end{figure}

The Hall mobility and density as functions of gate voltage,  allowing the effect of the gate to equilibrate, were shown in the Fig.~1 of the main text. Fig.~\ref{Fig:nmu} shows the Hall mobility and density as functions of gate voltage when the gate is swept upward, rather than being allowed to fully equilibrate.

\begin{figure}[ht]
	\includegraphics[width=0.35\textwidth]{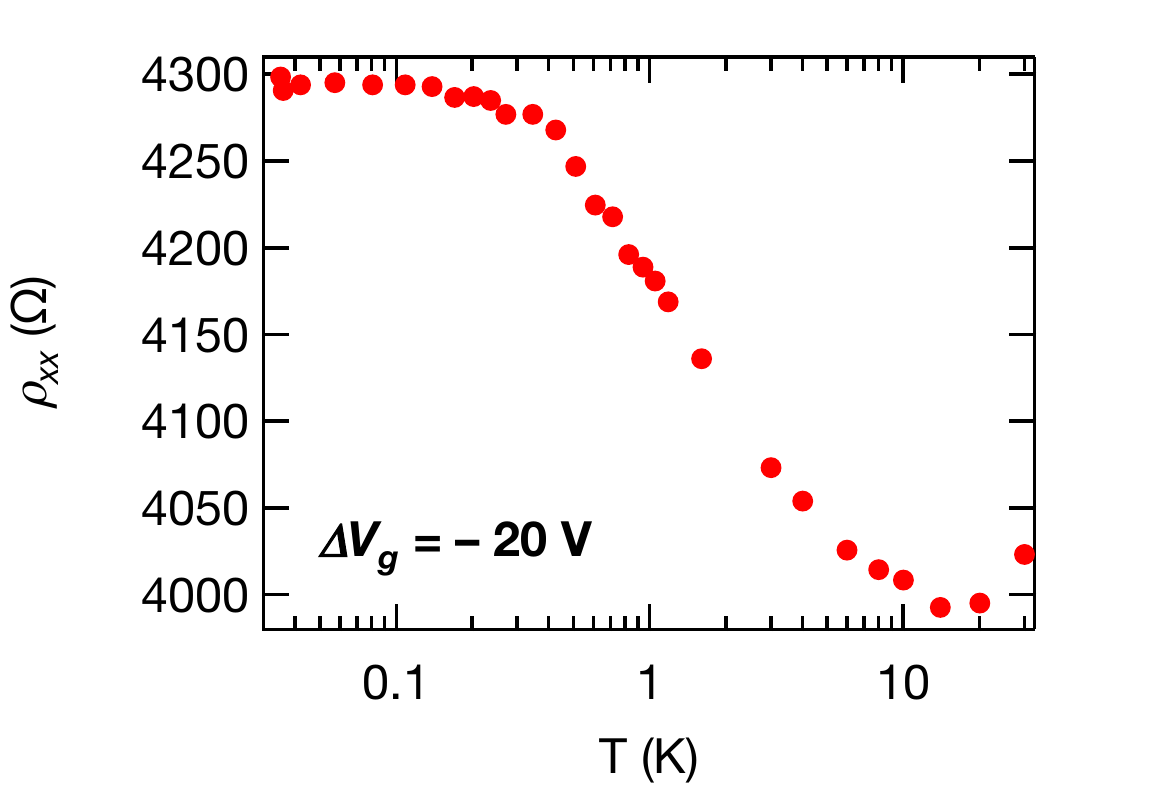}
	\caption{The resistivity at $V_g-V_\text{min}=-20$~V is shown with respect to temperature on a log scale. At this gate voltage, the system is substantially hole-doped, having density $n=3.3\times10^12~\text{cm}^{-2}$. Here, at intermediate temperatures, $\rho_{xx}$ exhibits the logarithmic temperature dependence characteristic of electron-electron interactions.}
	\label{Fig:logT}
\end{figure}

At substantial hole doping, the resistivity weakly decreases with increasing temperature. As shown in Fig.~\ref{Fig:logT}, the resistivity varies logarithmically in temperature over an intermediate temperature range. Negative logarithmic corrections to the resistivity in temperature are typically associated with weak localization, but since the magnetoconductance reveals pronounced weak anti-localization (which is associated with positive logarithmic temperature corrections), we instead connect the temperature dependence with electron-electron interactions. In accordance with our results, previous calculations have affirmed that the weak anti-localization effect should dominate in determining the magnetoconductance while electron-electron interactions should dominate in determining the temperature dependence~\cite{Lu2014a}.

\begin{figure}[ht]
	\includegraphics[width=0.45\textwidth]{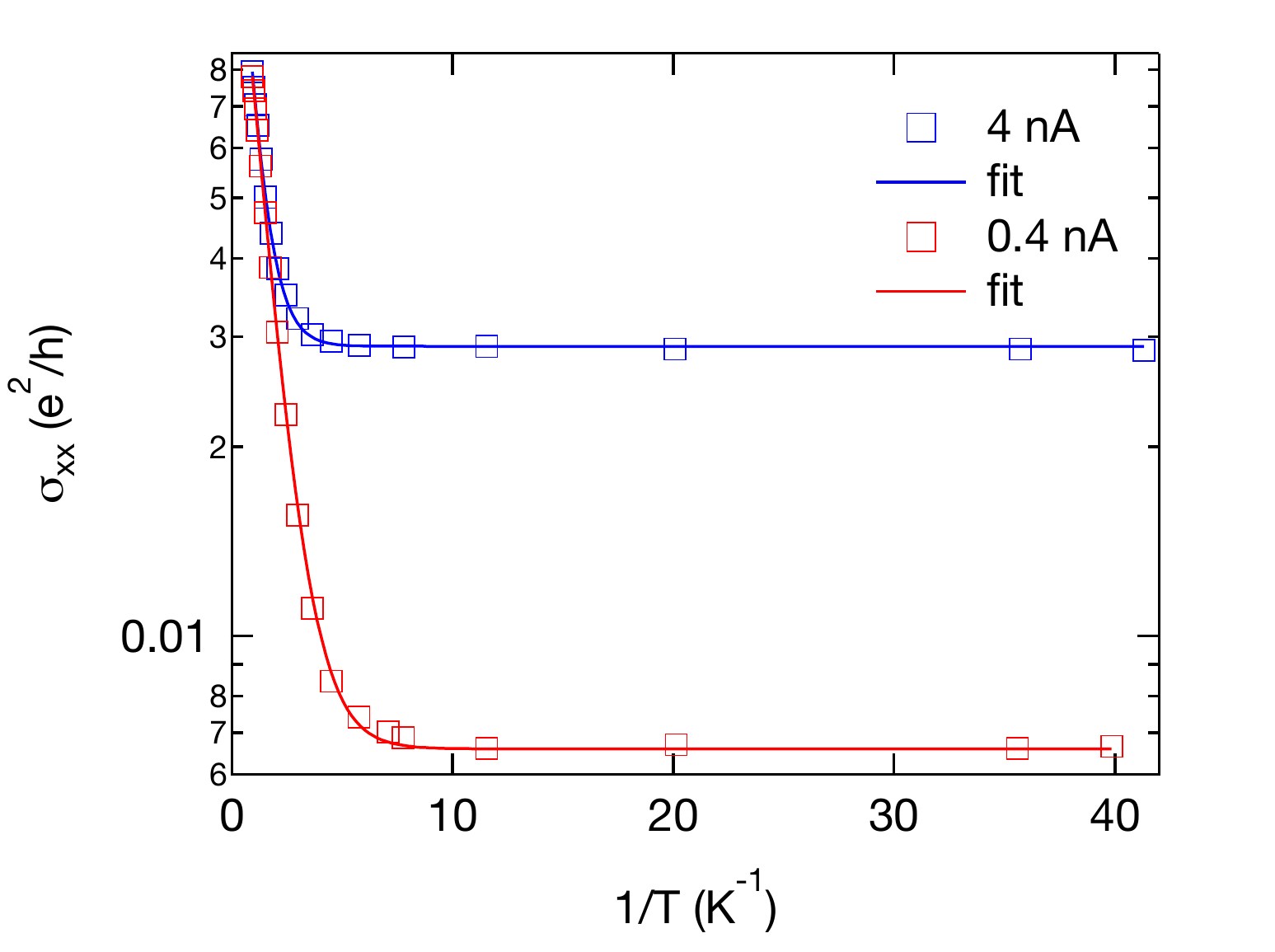}
	\caption{Temperature dependence of conductivity at two different current bias values. The conductivity versus temperature at $V_g=V_\text{min}$ is shown on an Arrhenius plot at current biases (blue squares) 4~nA and (red circles) 0.4~nA. Fits (solid lines) to an Arrhenius law plus a constant offset indicate a thermally activated scale of 0.98~K for both current bias values. At low temperatures as read on a thermometer attached to the low temperature stage of the cryostat, the conductivity saturates due to Joule heating, as manifested by the different saturation conductivities: 0.029 $e^2/h$ at 4~nA bias current, versus 0.0066 $e^2/h$ at 0.4~nA.}
	\label{Fig:ArrCur}
\end{figure}

\begin{figure}[ht]
	\includegraphics[width=0.45\textwidth]{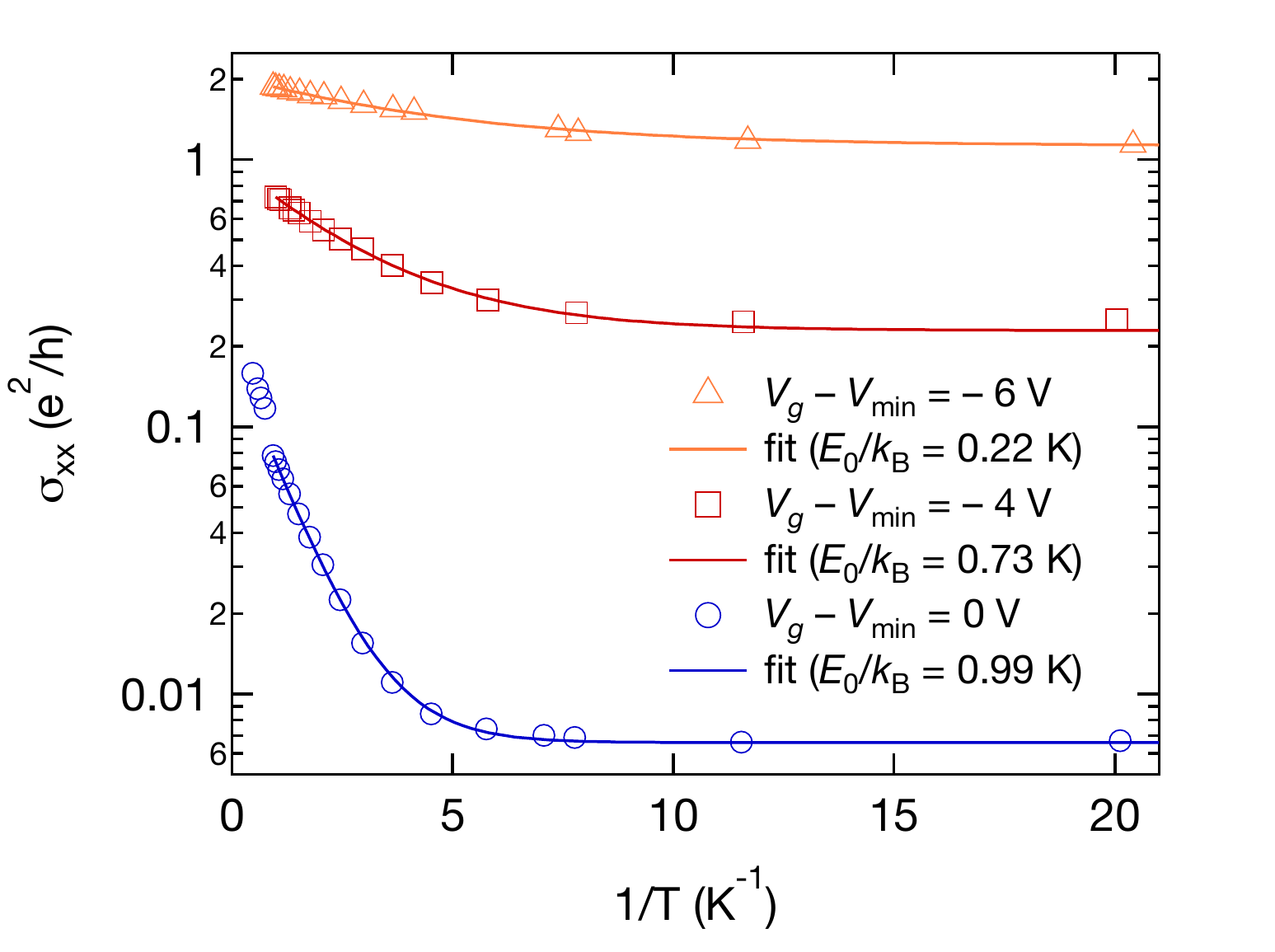}
	\caption{Temperature dependence of conductivity at three different gate voltage values. The conductivity versus temperature is shown on an Arrhenius plot at (blue circles) $V_g=V_\text{min}$, (red squares) $V_g-V_\text{min} = -4$~V, and (orange triangles) $V_g-V_\text{min} = -6$~V. Fits (solid lines) are to an Arrhenius law plus a constant offset. The extracted activation energies are indicated.}
	\label{Fig:ArrGate}
\end{figure}

Closer to the charge neutrality point, the temperature dependence of the resistivity becomes stronger. The conductivity at $V_g=V_\text{min}$ is fit by an Arrhenius law with a constant offset $\sigma_{xx} = \sigma_0 + \sigma_1 \exp(-\Delta_\text{Arr}/k_B T)$. The Arrhenius behavior indicates that the system is gapped. Since the offset $\sigma_0$ depends heavily on the bias current (Fig.~\ref{Fig:ArrCur}), we associate it with a divergence of the electron temperature from the thermometer temperature due to Joule heating. The temperature dependence at different gate voltages is shown on an Arrhenius plot in Fig.~\ref{Fig:ArrGate}, demonstrating weakening temperature dependence at $V_g$ further from $V_\text{min}$.

\section{Additional magnetoconductance data}

\begin{figure*}[ht]
	\includegraphics[width=0.95\textwidth]{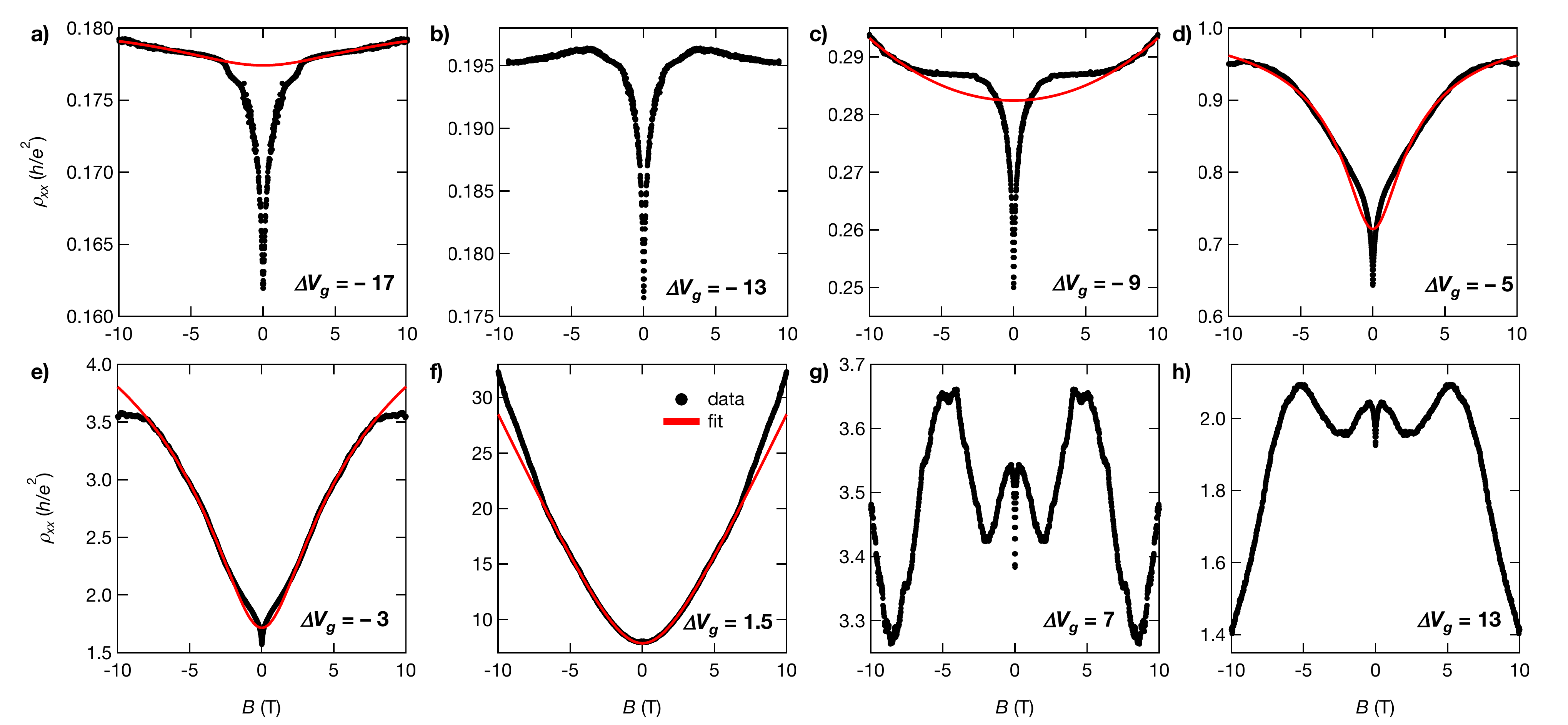}
	\caption{Classical magnetoconductance. a-h) Longitudinal resistivity is shown as a function of field for applied fields $B\leq 10\text{ T}$ at different gate voltages. The data (black) are symmetrized between positive and negative applied fields. Fits to a phenomenological model (red) for the classical magnetoconductance are shown for gate voltages at which a good fit was obtained.}
	\label{Fig:classMC}
\end{figure*}

\begin{figure*}[ht]
	\includegraphics[width=0.95\textwidth]{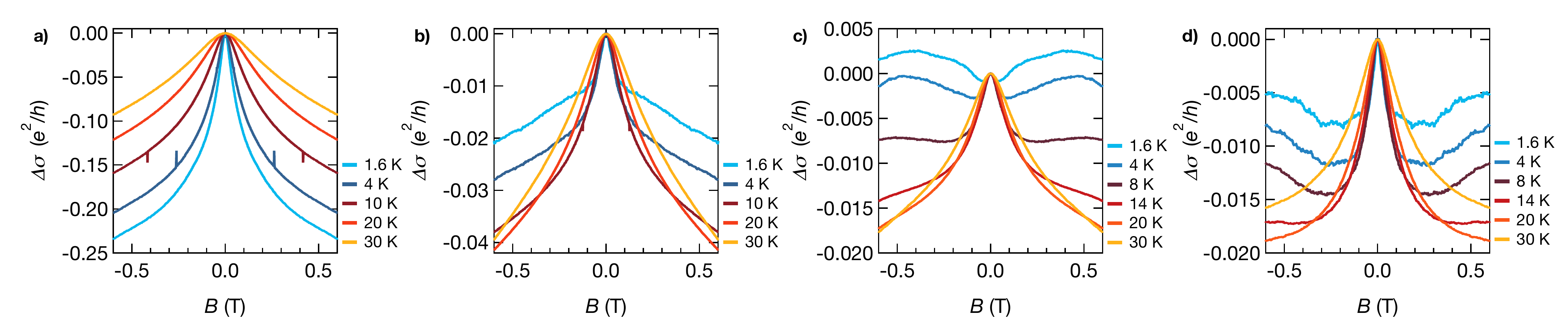}
	\caption{Temperature dependence of the WAL peak. a-d) The differential longitudinal conductivity, $\Delta\sigma_{xx}(T,H) = \sigma_{xx}(T,H)-\sigma_{xx}(T,0)$ at temperatures between $1.6\text{ K}$ and $30\text{ K}$ is shown at a) $V_g-V_\text{min} = -6\text{ V}$, b) $-2\text{ V}$, c) $2\text{ V}$, d) $8\text{ V}$.}
	\label{Fig:MCvsT}
\end{figure*}

The resistivity of the device as a function of magnetic field up to $\abs{B}=10$~T is shown in Fig.~\ref{Fig:classMC}, revealing the classical contribution to the magnetoconductance. Data is fit to a phenomenological model for the classical magnetoconductance in two-dimensional materials. At $V_g =24$~V and 30~V, when the negative Hall slope indicates an n-type carrier, the slope of the magnetoconductance repeatedly switches sign with increasing magnetic field. These are not Shubnikov--de Haas oscillations, but we lack an explanation. Supplementary magnetoconductance data at low magnetic field and various temperatures are shown in Fig.~\ref{Fig:MCvsT}, including data at $V_g-V_\text{min}=-6$~V, $-2$~V, 2~V, and 8~V.  

\begin{figure*}[ht]
	\includegraphics[width=0.75\textwidth]{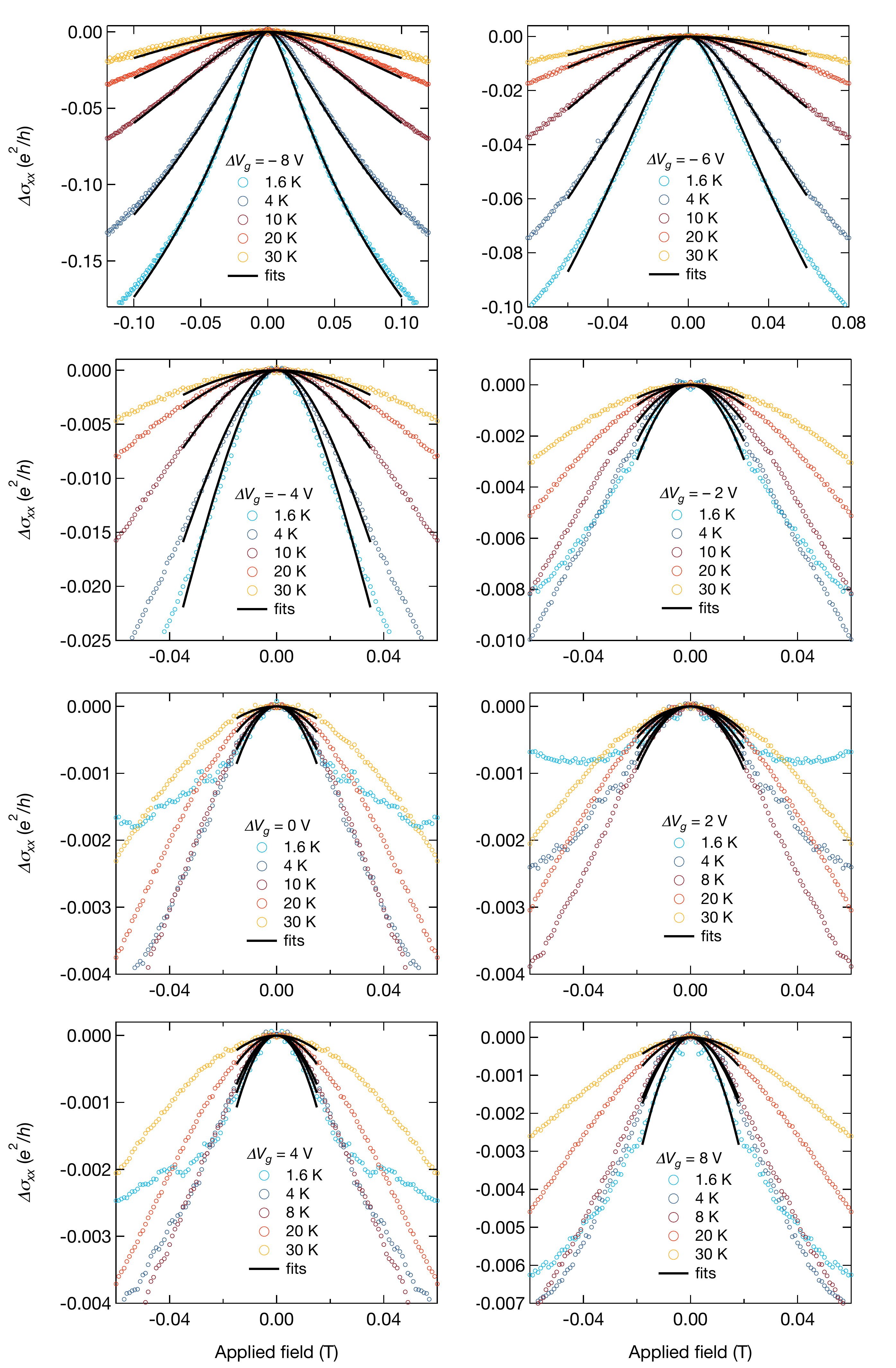}
	\caption{Extracting the coherence length $l_\phi$. The differential longitudinal conductivity, $\Delta\sigma_{xx}(T,H) = \sigma_{xx}(T,H)-\sigma_{xx}(T,0)$ at temperatures between $1.6\text{ K}$ and $30\text{ K}$ is shown at various gate voltages (colored circles). The HLN formula with only a WAL term is fit to the data (black lines) in order to extract the coherence length $l_\phi$. A limited range in magnetic field is used in the fits to avoid contribution from the crossover to WL. The contribution of the classical magnetoconductance has not been subtracted.}
	\label{Fig:MCfits}
\end{figure*}

Fig.~\ref{Fig:MCfits} shows the data and fits used to extract the coherence lengths $l_\phi$ shown in Fig.~4 of the main text. Data and fits at some temperatures are omitted for clarity.

\section{Percolative transport}

One might argue that, due to inhomogeneities, electronic transport in the charge puddle regime (i.e. when $V_g-V_\text{min}\approx 0$) may be mediated by percolation, so that transport is relegated to thin conductive tendrils. Since the local conductivity in these tendrils would be higher than the nominal conductivity from our measurements (which yield an average of the conductivity throughout the device), could we not be satisfying the Ioffe-Regel criterion $\sigma_{xx} \gtrsim e^2/h$ even at $V_g = V_\text{min}$? While we acknowledge that sub-threshold percolation is likely present, percolation fails to explain our observation of a WAL coherence peak when the measured conductivity falls beneath $e^2/h$. The electronic states in a material become one-dimensional past the percolation threshold. According to localization theory, when the conductance of a one-dimensional system $G$ is substantially smaller than $e^2/h$, the wavefunctions cannot be itinerant and therefore must be strongly localized.

\clearpage

%
